# Nanophotonic chiral sensing: How does it actually work?

Steffen Both[1], Martin Schäferling[1], Florian Sterl[1], Egor Muljarov[2], Harald Giessen[1], and Thomas Weiss[1]

[1]4th Physics Institute and Research Center SCoPE, University of Stuttgart, Pfaffenwaldring 57, 70569 Stuttgart, Germany
[2]Cardiff University, School of Physics and Astronomy, The Parade, CF24 3AA, Cardiff, UK

## Abstract

Nanophotonic chiral sensing has recently attracted a lot of attention. The idea is to exploit the strong light-matter interaction in nanophotonic resonators to determine the concentration of chiral molecules at ultra-low thresholds, which is highly attractive for numerous applications in life science and chemistry. However, a thorough understanding of the underlying interactions is still missing. The theoretical description relies on either simple approximations or on purely numerical approaches. We close this gap and present a general theory of chiral light-matter interactions in arbitrary resonators. Our theory describes the chiral interaction as a perturbation of the resonator modes, also known as resonant states or quasi-normal modes. We observe two dominant contributions: A chirality-induced resonance shift and changes in the modes excitation and emission efficiencies. Our theory brings new and deep insights for tailoring and enhancing chiral light-matter interactions. Furthermore, it allows to predict spectra much more efficiently in comparison to conventional approaches. This is particularly true as chiral interactions are inherently weak and therefore perturbation theory fits extremely well for this problem.

## Introduction

The term "chirality" refers to objects that cannot be superimposed with their mirror image[1]. These two so-called enantiomorphs (or, in case of molecules, enantiomers) differ only in their handedness, which can be left or right. What sounds like a purely mathematical concept has in fact a huge impact, as life itself is chiral[2,3]. The outcome of most biochemical interactions, where chiral biomolecules shake hands, strongly depends on the mutual handedness of the reactants. In extreme examples, the handedness of a molecule makes the difference between a drug and a toxin[4,5]. Therefore, detecting the handedness of molecules is of crucial interest for countless applications in life science and chemistry, as well as for the pharmaceutical industry[6].

Conventional detection schemes rely on the fact that the interaction of chiral media with light can differ among the two circular polarizations and depends on the handedness of the enantiomers. Assuming a homogeneous and isotropic medium, this interaction is governed by the chiral constitutive equations (below provided in Gaussian units)[7,8]:

$$\begin{aligned} \mathbf{D} &= \varepsilon \mathbf{E} - i\kappa \mathbf{H}, \\ \mathbf{B} &= \mu \mathbf{H} + i\kappa \mathbf{E}. \end{aligned} \quad (1)$$

Here, the permittivity $\varepsilon$ and the permeability $\mu$ represent the "nonchiral" properties of the medium, and the Pasteur parameter $\kappa$ quantifies its chirality. Opposite handedness of the medium results in an opposite sign of $\kappa$. A nonzero real part of $\kappa$ induces a difference in the phase velocities of left and right-handed circularly polarized light, i.e., circular birefringence, while a nonzero imaginary part induces a difference in their absorption. Measuring this absorption difference – denoted as the circular-dichroism (CD) signal – is the standard method for optically characterizing the chirality of a medium.

Since chiral light-matter interactions are typically extremely weak (at optical frequencies, natural materials have $\kappa \ll 1$), this detection can be very challenging, especially when only tiny amounts of substances are involved. Overcoming this limitation would be highly attractive for numerous applications. A promising approach consists in the use of nanophotonic resonators to boost the chiral light-matter interactions. For the sensing of "nonchiral" material properties, this is already a well-established technique. Applications include the ultra-sensitive detection of biomolecules[9–15], gases[16,17], and much more[18–20]. In the past decade, a lot of work, both experimental[21–33] and theoretical[30,34–58], has been carried out to utilize the benefits of this technique for the detecion of chiral substances. Different resonator designs have been investigated, ranging from plasmonic antennas[21–28,38,42] to dielectric nanostructures[29,49,52,54,55] or combinations[50] to so-called "helicity-preserving" cavities[53,56–58]. Another promising route is using structures that exhibit resonances in the ultraviolet region[31–33], where natural molecules have particularly large $\kappa$ values. Comprehensive overviews can be found in corresponding review articles[59–64].

The basic principle of nanophotonic chiral sensing is illustrated in Fig. 1a,b: The starting point is a nanophotonic res-onator that can be brought into contact with which the chiral medium. The resonator itself may also be chiral (e.g., due to its shape), but does not have to be. As an example, we depict an array of Ω-shaped plasmonic nanoantennas[65–68] with the chiral medium in their centers. First, the CD spectrum of the resonator without the chiral medium is measured (gray), to serve as a reference. Then, the resonator is brought into contact with the chiral medium and the CD spectrum is measured again (light blue). Note that for visualization, the spectral changes in the plot are dramatically exaggerated. The difference between both spectra (we denote it as ΔCD) contains information about the handedness of the medium. Due to the enhanced light-matter interaction taking place in nanophotonic resonators, the ΔCD signal is typically orders of magnitude larger than the signal that would be obtained from the chiral medium alone.

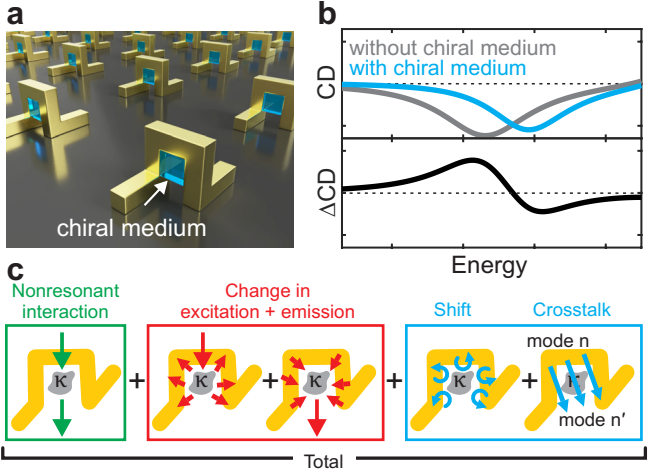

**Fig. 1. Principle of nanophotonic chiral sensing and underlying contributions. a** A nanophotonic resonator (example depicts an array of Ω-shaped plasmonic antennas) with a chiral medium in its center. **b** The circular-dichroism (CD) spectrum is measured without (gray) and with (light blue) chiral medium. The change in the spectrum (ΔCD) contains information about the chirality of the medium. **c** As shown in this work, the total interaction can be explained as a combination of five different contributions: A nonresonant interaction, changes in excitation and emission efficiencies of the modes, modal resonance shifts, and intermodal crosstalk. Note that due to similarities that will be discussed later, the contributions are categorized into three different groups.

An important experimental detail in the above procedure is not to use the plain resonator as reference, but rather the resonator covered with a so-called racemic mixture[28,29] (1:1 mixture of left-handed and right-handed enantiomers, which is optically achiral) at the positions where the chiral medium is supposed to be placed later. This ensures that only $\kappa$ varies between both measurements, while the other material parameters are constant. Note that there also exist variations of the procedure that work without the need to use a racemic mixture[21,46]; however, also in these cases, the key lies in tracing the change of an optical signal induced by the interaction of the resonator with the chiral medium.

For the case of a single chiral molecules, the above interaction is well understood[36,37]; however, in practice, one typically does not deal with single chiral molecules, but with chiral media (i.e., a solution or a layer of many molecules). In this case, describing the interaction with a resonator is more sophisticated. The description relies on either simple approximations or on purely numerical approaches: The intuitive method[21,30,39,41,44,47,48] consists in evaluating the optical chirality[35] of the resonator's near-field. This allows for predicting the power absorbed in the chiral medium. However, while being very illustrative, this approach has severe limitations[29,62,69]: First, it neglects any influence of the real part of $\kappa$, which – albeit it would not contribute to the CD of the chiral medium located outside a resonator – is known to strongly contribute to the CD of the combined system[29,69]. Second, it neglects the back action from the chiral medium onto the fields of the resonator, known as induced CD[37,69,70]. The rigorous method[27,29,46,51,57,69–73] consists in directly including the chiral medium into numerical calculations via Eq. (1). However, while this approach accounts for all electromagnetic effects, it provides rather limited insights into the interaction. As an alternative to numerical calculations, in some cases, the interaction can be described analytically via Mie theory[74], or semianalytically via a simple closed-form expression[49]. The former is, however, only applicable for systems with spherical or ellipsoidal symmetry, while the latter only works for resonators that can be treated as an effective medium.

We close the existing gap and present a general theory of chiral light-matter interactions in arbitrary resonators. Our theory retains the rigorousness of the numerical calculations, while at the same time providing a deep intuitive insight. It describes the chiral interaction as a perturbation of the modes of the resonator, also known as resonant states[75–80] or quasi-normal modes[81–86]. We show that the *entire chiral light-matter interaction* can be explained as a combination of five different contributions (illustrated in Fig. 1c): a nonresonant interaction, changes in the excitation and emission efficiencies of the modes, modal resonance shifts, and intermodal crosstalk. Note that the contributions can be organized into three different categories, due to similarities that will be discussed later. We quantify the impact of these contributions in different sensor geometries. Furthermore, we show that – contrary to common expectation – resonance shifts are often not the dominating source of signal.

Describing nanophotonic resonators via their modes is a highly efficient approach that is experiencing rapidly increasing recognition in the community[12,20,75–98]. Our derivations are based on previous works from the field: In Ref. [76], a rigorous electrodynamic perturbation method was developed for predicting modal changes in systems containing bi-anisotropic materials. References [77, 80, 95] demonstrate how to construct the optical scattering matrix of a resonator from its modes. Based on these works, we have derived a simple expression for the change of the optical scattering matrix under chiral material perturbations.

## Results

**Theory.** In the following, we give a brief summary of the theory. A detailed derivation can be found in the Supporting Information. We start from Maxwell's equations for an optical resonator without any chiral medium. For compactness of notations, we make use of the operator formalism introduced in Ref. [76]. Thus, Maxwell's equations read [Gaussian units, frequency domain, time dependence $\exp(-i\omega t)$, no external currents]:

$$\underbrace{\left[k\begin{pmatrix}\varepsilon(\mathbf{r};k) & 0 \\ 0 & \mu(\mathbf{r};k)\end{pmatrix} - \begin{pmatrix}0 & \nabla\times \\ \nabla\times & 0\end{pmatrix}\right]}_{\hat{\mathbb{M}}(k)} \underbrace{\begin{pmatrix}\mathbf{E}(\mathbf{r}) \\ i\mathbf{H}(\mathbf{r})\end{pmatrix}}_{\mathbb{F}} = 0. \quad (2)$$

Here, $k$ represents the vacuum wavenumber (for brevity of notations, we will consistently use $k = \omega/c$ instead of the frequency $\omega$); the permittivity $\varepsilon(\mathbf{r};k)$ and permeability $\mu(\mathbf{r};k)$ describe the spatial material distribution of the resonator, and $\mathbf{E}$ and $\mathbf{H}$ denote the electric and magnetic field, respectively. By combining the fields into a six-dimensional supervector $\mathbb{F}$, and summarizing all other quantities into the Maxwell operator $\hat{\mathbb{M}}$, Maxwell's equations become a simple one-line equation: $\hat{\mathbb{M}}\mathbb{F} = 0$. The modes of the resonator are defined as the solutions of this equation that simultaneously satisfy outgoing boundary conditions. One finds[76]

$$\hat{\mathbb{M}}(k_n)\mathbb{F}_n = 0, \quad (3)$$

where $n$ is an index that labels the modes, $\mathbb{F}_n$ denote the modal field distributions, and $k_n$ are the corresponding wavenumber eigenvalues. In general, $k_n$ are complex, with $\text{Re}(k_n)$ representing the resonance wavenumbers of the modes and $-2\,\text{Im}(k_n)$ denoting their linewidths. Since nanophotonic resonators are typically open systems[82] (they leak energy to the environment), their modes are generally referred to as resonant states or quasi-normal modes. For the sake of brevity, we will continue to use the intuitive term modes. Note that Eq. (3) defines $\mathbb{F}_n$ only up to an arbitrary scalar factor. In order to be applicable for an expansion, they have to be normalized. Valid normalization schemes can be found in Refs. [75, 76, 80, 82, 83, 93], and references therein.

A very convenient formalism to summarize the interaction of a resonator with incident light is via the optical scattering matrix $S$[77,80,95,98]. The idea is as follows: The resonator has different so-called incoming channels $\mathbf{N}$, via which it can be excited, and different so-called outgoing channels $\mathbf{M}$, via which it can radiate light[80,99]. Here, the vectors $\mathbf{N}$ and $\mathbf{M}$ each represent a set of quantum numbers that specify details about the respective channel (e.g., polarization, propagation direction,...). Each element $S_{\mathbf{MN}}$ of the scattering matrix represents the transmission (or reflection) amplitude from one particular input channel $\mathbf{N}$ into one particular output channel $\mathbf{M}$. The scattering matrix can be calculated from the modes as[77,80,95]

$$S_{\mathbf{MN}} = S_{\mathbf{MN}}^{\text{bg}} + \sum_n \frac{a_{n,\mathbf{M}} b_{n,\mathbf{N}}}{k - k_n}. \quad (4)$$

Here, $S_{\mathbf{MN}}^{\text{bg}}$ corresponds to a nonresonant background term, while $a_{n,\mathbf{M}}$ and $b_{n,\mathbf{N}}$ represent the emission and excitation coefficients, respectively, of the modes $\mathbb{F}_n$ (expressions provided in the Supporting Information). Furthermore, $k$ denotes the wavenumber at which the resonator is excited. Note that $S_{\mathbf{MN}}^{\text{bg}}$, $a_{n,\mathbf{M}}$, and $b_{n,\mathbf{N}}$ are considered here as $k$ dependent. We want to remark that there exist several alternative representations[77,80,95,98] of Eq. (4), which differ in the definition of the background term and the coefficients.

Now, let us assume the resonator is perturbed by locally inserting a chiral medium. Mathematically, this can be accounted for by changing the operator $\hat{\mathbb{M}}$ that describes the resonator to $\hat{\mathbb{M}} + \delta\hat{\mathbb{M}}$, with the perturbation operator $\delta\hat{\mathbb{M}}$, which is defined as[76]

$$\delta\hat{\mathbb{M}} = \begin{cases} k\begin{pmatrix}0 & -\kappa \\ -\kappa & 0\end{pmatrix} & \text{inside volume } V_c, \\ 0 & \text{outside.} \end{cases} \quad (5)$$

Here, $V_c$ represents the volume in which the chiral medium is inserted, and $\kappa$ is the Pasteur parameter. The above $\delta\hat{\mathbb{M}}$ corresponds to the most relevant scenario, where one transitions from a racemic mixture to a chiral medium, such that only $\kappa$ varies and the "nonchiral" material parameters $\varepsilon$ and $\mu$ stay constant. It is straightforward though, to extend $\delta\hat{\mathbb{M}}$ for changes in $\varepsilon$ and $\mu$ as well. Furthermore, instead of considering a scalar $\kappa$, it is also possible to include bi-anisotropic contributions, originating, e.g., from molecular alignment effects[27,73]. The most general $\delta\hat{\mathbb{M}}$ can be found in the the Supporting Information.

As a consequence of the perturbation, the scattering matrix changes from $S$ to $S + \delta S$, where $\delta S$ denotes the change. In the case of chiral media, one can safely assume that the perturbation is small compared to the unperturbed material parameters. Therefore, one can apply a first-order perturbation theory. After some derivations (see Supporting Information), we obtain the change of the scattering matrix as

$$\delta S = \delta S^{\text{nr}} + \delta S^{\text{ex}} + \delta S^{\text{em}} + \delta S^{\text{shift}} + \delta S^{\text{cross}}, \quad (6)$$

which contains five contributions defined as:

$$\delta S_{\mathbf{MN}}^{\text{nr}} = \int_{V_c} ik\kappa \left(\mathbf{E}_{\mathbf{M}}^{\text{R}} \cdot \mathbf{H}_{\mathbf{N}} + \mathbf{H}_{\mathbf{M}}^{\text{R}} \cdot \mathbf{E}_{\mathbf{N}}\right) dV, \quad (7)$$

$$\delta S_{\mathbf{MN}}^{\text{ex}} = -\sum_n \frac{a_{n,\mathbf{M}} \int_{V_c} ik\kappa \left(\mathbf{E}_n^{\text{R}} \cdot \mathbf{H}_{\mathbf{N}} + \mathbf{H}_n^{\text{R}} \cdot \mathbf{E}_{\mathbf{N}}\right) dV}{k - k_n}, \quad (8)$$

$$\delta S_{\mathbf{MN}}^{\text{em}} = -\sum_n \frac{b_{n,\mathbf{N}} \int_{V_c} ik\kappa \left(\mathbf{E}_{\mathbf{M}}^{\text{R}} \cdot \mathbf{H}_n + \mathbf{H}_{\mathbf{M}}^{\text{R}} \cdot \mathbf{E}_n\right) dV}{k - k_n}, \quad (9)$$

$$\delta S_{\mathbf{MN}}^{\text{shift}} = \sum_n \frac{a_{n,\mathbf{M}} b_{n,\mathbf{N}} \int_{V_c} ik\kappa \left(\mathbf{E}_n^{\text{R}} \cdot \mathbf{H}_n + \mathbf{H}_n^{\text{R}} \cdot \mathbf{E}_n\right) dV}{(k - k_n)^2}, \quad (10)$$

$$\delta S_{\mathbf{MN}}^{\text{cross}} = \sum_{n \neq n'} \frac{a_{n,\mathbf{M}} b_{n',\mathbf{N}} \int_{V_c} ik\kappa \left(\mathbf{E}_n^R \cdot \mathbf{H}_{n'} + \mathbf{H}_n^R \cdot \mathbf{E}_{n'}\right) dV}{(k - k_n)(k - k_{n'})}.$$
(11)

Here, $\mathbf{E_M}, \mathbf{H_M}$, and $\mathbf{E_N}, \mathbf{H_N}$ denote the background fields belonging to the $\mathbf{M}$-th outgoing and the $\mathbf{N}$-th incoming channel, respectively (for details, see Supporting Information), while $\mathbf{E}_n, \mathbf{H}_n$ represent the fields of mode $n$. The superscript R indicates reciprocal conjugation[80], which is included for the sake of generality and is, e.g., needed when dealing with periodic systems under oblique incidence angles[96]. In most practically relevant cases (for details, see Refs. [76, 80, 95]), one trivially obtains $\mathbf{E}^R = \mathbf{E}$ and $\mathbf{H}^R = \mathbf{H}$.

Equations (6) to (11) are the main result of this work. They allow to predict the response of a resonator to chiral material changes via simple overlap integrals of the unperturbed fields over the region of the perturbation. We want to remark that, although we are only interested in $\kappa$ changes, the above equations can be easily extended to account for changes in $\varepsilon$, $\mu$, and the bi-anisotropic parameters as well. The general expressions are provided in the Supporting Information.

As shown above, the total change $\delta S$ is composed of five contributions. Every contribution describes the effect of a different perturbation-induced physical process on the scattering matrix. The first one, $\delta S^{\text{nr}}$, contains an overlap integral between incoming and outgoing background fields and represents a nonresonant interaction. The second, $\delta S^{\text{ex}}$, and third contribution, $\delta S^{\text{em}}$, consist of overlap integrals of the modes with the background fields, which describe changes in the excitation and emission efficiencies, respectively. The relevance of this overlap has been predicted by us in a previous numerical study[70]. The above equations now rigorously prove this prediction. The fourth contribution, $\delta S^{\text{shift}}$, contains an overlap integral of the modes with themselves, which is associated with a shift of their wavenumber eigenvalues. The fifth contribution, $\delta S^{\text{cross}}$, comprises an overlap integral between different modes, which describes perturbation-induced crosstalk. Note that the shift and the crosstalk contributions are mathematically very similar. Therefore, we categorized them into a common group of effects in the illustrations of Fig. 1. We will later investigate the significance of these contributions in different example systems.

Let us now have a closer look at the shift contribution. As already indicated, $\delta S^{\text{shift}}$ denotes the response of the scattering matrix to resonance shifts. The shift of the wavenumber eigenvalue of an individual mode is given as

$$\Delta k_n = \int_{V_c} \underbrace{ik_n\kappa \left(\mathbf{E}_n^R \cdot \mathbf{H}_n + \mathbf{H}_n^R \cdot \mathbf{E}_n\right)}_{\text{``}\Delta k_n \text{ per volume''}} dV.$$
(12)

This is the chiral analog of the eigenvalue shift that is well-known from previous works on "nonchiral" sensing for $\varepsilon$ and $\mu$ changes[20,75,79,81,82,96,97], where it is considered to be the crucial quantity to maximize the sensitivity. The term under the integral can be interpreted as a "shift per volume" density. In general, $\Delta k_n$ is complex, with $\text{Re}(\Delta k_n)$ corresponding to a change in the resonance wavenumber and $-2\,\text{Im}(\Delta k_n)$ representing a change in the linewidth.

The above equation reveals an interesting connection to the optical chirality introduced by Tang and Cohen[35]: Let us assume a typical chiral medium with $|\text{Re}(\kappa)| \gg |\text{Im}(\kappa)|$, a resonator with low losses, i.e., $|\text{Re}(k_n)| \gg |\text{Im}(k_n)|$, and furthermore restrict the considerations to a scenario where $\mathbf{E}^R = \mathbf{E}$ and $\mathbf{H}^R = \mathbf{H}$. In this case, one finds that the change of the resonance wavenumber $\text{Re}(\Delta k_n)$ is proportional to the integral over $\text{Im}(\mathbf{E}_n \cdot \mathbf{H}_n)$. This quantity is closely related to the optical chirality. Therefore, as a rule of thumb, one can deduce that systems optimized for strong chiral modes are also sensitive for resonance wavenumber changes.

**Example 1: Rod antennas.** In the following, we will apply our theory to different examples of nanoresonator systems. As a first example, we consider one of the most frequently used structures in nanophotonic sensing: an array of plasmonic rod antennas. The geometry is depicted in Fig. 2a. The dimensions are chosen to achieve a resonance in the near-infrared spectral range (for details, see Methods section). The antennas consist of gold and are surrounded by water. Chiral media patches are placed at the ends of each antenna, i.e., in the regions where the strongest near-fields occur. The chiral medium is accounted for with a Pasteur parameter of $\kappa = (1 + 0.01i) \times 10^{-4}$. This value is deliberately chosen such that it exhibits a large but still realistic magnitude[29,100] and contains a typical ratio between real and imaginary part[29,70].

Figure 2b shows the spectral response of the antennas around their fundamental plasmonic mode, which is found at an energy of $1146.1 - 51.3i$ meV. As a representative quantity, we plot the $\Delta$CD signal. The corresponding matrices $S$ and $\delta S$ with all their components can be found in the Supplementary Figs. 1 and 2. Details on how the $\Delta$CD signal is obtained from $S$ and $\delta S$ are provided in the Supplementary Information. The matrices were calculated from inserting the fundamental mode into Eqs. (4) and (6). To improve the accuracy of $S$, an additional cubic fit was used as background to account for the influence of higher-order modes (for details, see Methods section).

The top panel in Fig. 2b depicts the total $\Delta$CD signal. The line denotes the result of the modal theory, while the dots have been obtained by exact full-wave calculations (for details, see Methods section) and are plotted for comparison. It can be seen that there is excellent agreement. The total $\Delta$CD signal exhibits a Lorentzian line shape, with its highest absolute value being located at the resonance energy of the mode (indicated by a vertical dashed line). The lower panels display the different contributions. For compactness, we have summed up the change in the excitation and emission efficiency contributions to one curve. Furthermore, there is no crosstalk curve, since only one mode is considered. Note that the curve for the nonresonant interaction shows a zero crossing at exactly the resonance energy of the mode. This might strike as a mistake, since $\delta S^{\text{nr}}$ does not contain any modal dependence [cf.

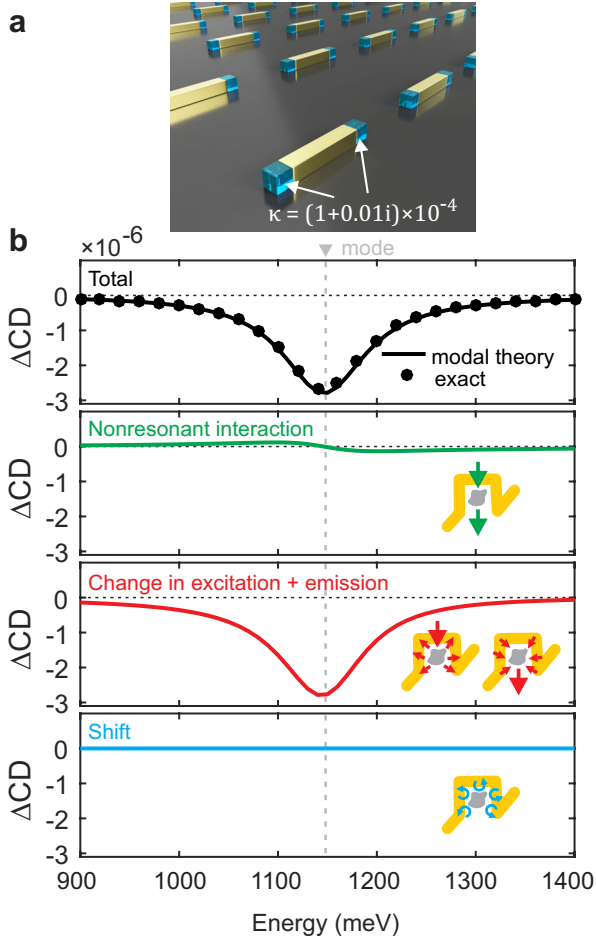

**Fig. 2. Optical response of plasmonic rod antennas. a** We consider an array of rod antennas with chiral media patches located at each end. **b** Resulting ΔCD spectrum. The top plot denotes the total signal (line: calculated with our modal theory; dots: exact full-wave calculations). The individual contributions are depicted below. The system is dominated by the changes in the excitation and emission efficiencies, while the shift contribution is exactly zero.

Eq.(7)]. There is, however, a trivial explanation: As already mentioned, we do not look at the contributions of $\delta S$, but at their impacts on the ΔCD signal. These impacts contain an additional modulation by the unperturbed scattering matrix $S$ (details, see derivation of ΔCD in the Supporting information) and this matrix has a modal dependence.

The results in Fig. 2b paint a very clear picture: The system is dominated by the change in the excitation and emission efficiencies of the mode. The nonresonant interaction contribution is very small and practically irrelevant. The shift contribution is not only small but turns out to be strictly zero. This result is quite surprising, since resonance shifts are widely believed to be the driving mechanism behind nanophotonic chiral sensing[21,25,27,38].

In order to understand why the shift contribution is zero, it is instructive to have a closer look at the mode. Its electric field is visualized in Fig. 3a. By applying Eq. (12), one can derive the corresponding "shift per volume" density. The result is plotted in Fig. 3b. Note that for consistency with the spectra, we use units of energy ("$\Delta E_n$ per volume") instead units of wavevector ("$\Delta k_n$ per volume"). The shift of the energy eigenvalue $\Delta E_n$ ($=c\hbar\Delta k_n$) is obtained by integrating the "shift per volume" density over the volume of the perturbation (the chiral media patches are displayed in Fig. 3b as bluish cubes). From the plot, it is obvious that the integral vanishes: First of all, the "shift per volume" is very weak inside the region of the patches. Second, and more importantly, positive (red) and negative (blue) contributions are occurring symmetrically such that they cancel out each other. It can be easily deduced from the plot that this symmetry argument does not only apply when the chiral medium is positioned at the ends of the antenna, but also holds for other distributions, e.g., when the medium would completely surround the antenna.

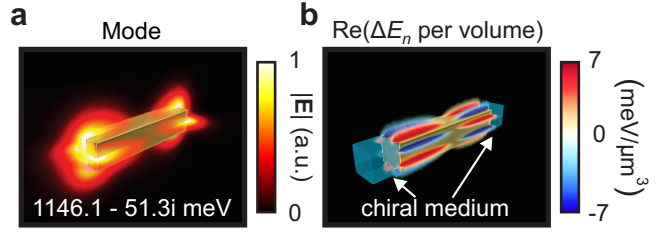

**Fig. 3. Details on the resonance shift in the rod antennas. a** Fundamental plasmonic mode. **b** "Shift per volume" of that mode. From the figure, it is obvious that inside the volume of the chiral media patches (displayed in panel **b** as bluish cubes), the "shift per volume" is very weak, and furthermore, positive and negative contributions cancel out each other.

In fact, it is straightforward to prove that *any geometrically achiral sensor* will experience zero resonance shift. The argument is as follows: The electric field classifies as a vector (it flips its direction under parity inversion), while the magnetic field classifies as a pseudovector (it does not flip its direction under parity inversion)[101]. This makes the shift $\Delta E_n$ defined by Eq. (12) a pseudoscalar (it does change its sign under parity inversion), and therefore vanishing for geometrically achiral sensors. f This derivation suggests, however, that it can be possible to enforce a frequency shift in the rod antennas by breaking the achiral symmetry of the patch arrangement. An obvious choice consists in placing the chiral medium only in regions with a uniform sign of the "shift per volume" density[40]. To verify this, we have considered the rod antenna with the chiral medium distributed over the positive regions (red spots in Fig. 3b). The results (see Supplementary Figs. 3 and 4) confirm that this arrangement indeed provides a nonzero shift contribution. However, interestingly, it can only be observed in the channels of $\delta S$, but not in the ΔCD spectrum, since the signals from different channels cancel out each other. Furthermore, it is quite obvious that such a three-dimensional patch arrangement would be rather difficult to realize in practice.

**Example 2: Ω antennas.** As a second example, let us now investigate a system that is specifically designed to generate a strong shift. Before we discuss the structure, let us recapitulate the requirements dictated by Eq. (12): On the one hand, the system needs to feature a mode with strong collinear electric and magnetic fields in some region of space. On the other hand, the product between both fields should exhibit one predominant sign. Many systems discussed in the literature on chiral sensing[21,38,39,42,44,46] are already optimized for strong optical chirality and – due to the connection discussed under Eq. (12) – should intrinsically feature a strong shift. Representatively for such structures, we choose a very intuitive geometry: the Ω-shaped antennas depicted in Fig. 1a. They can be understood as follows: Each Ω consists of an upright standing split-ring resonator with two rod antennas attached to its feet. While split rings are known to feature strong dipolar magnetic fields, rod antennas do support strong dipolar electric fields. The particular arrangement of these components promises collinearity of the fields within the center of the Ω. Although they have never been utilized in the context of sensing so far, such Ω antennas are known to exhibit a strong chiroptical far-field response[65–68]. We consider again a periodic array of antennas. The materials are the same as in the previous example, and the dimensions are chosen comparably (for details, see Methods section). Note, however, that the fabrication of such three-dimensional structures is in general not an easy challenge (although there are approaches[67,68,102,103]).

Figure 4a displays the "shift per volume" density of the energetically lowest two excitable plasmonic modes. The results confirm what was intuitively expected: There is a hotspot with high uniform values in the center of the antenna. Note that for the given configuration, mode 1 exhibits a negative sign, while mode 2 exhibits a positive one. Let us now assume that a patch of chiral medium is positioned in the hotspot (for visualization, see Fig. 1a). Figure 4b displays the resulting energy eigenvalue shifts $\Delta E_n$ (both the real and imaginary parts) as a function of $\kappa$. On the x axis, $\kappa$ is varied as a multiple of $\kappa_0 = (1 + 0.01i) \times 10^{-4}$. The lines show the prediction of the modal theory, while the dots have been derived from exact full-wave calculations and are depicted for comparison. It is evident that there is an excellent agreement. As expected from the "shift per volume" plots, the energy eigenvalues are very sensitive to $\kappa$ changes. In agreement with the sign of the "shift per volume" density in the hotspot, mode 1 shows a negative slope in the Re($\Delta E_n$) plot, while mode 2 exhibits a positive one.

To investigate the impact of the $\Delta E_n$ shifts, we evaluate again the spectral response of the system. In order to improve the accuracy, we include one further mode in the calculation. This mode is found at an energy eigenvalue of $1750.1 - 25.4i$ meV and contributes to the spectrum via crosstalk. The calculation results are displayed in Fig. 5. As in the rod antenna example, we depict only the ΔCD signal, while the full matrices $S$ and $\delta S$ can be found in the Supplementary Figs. 5 and 6, respectively. All calculations are anal-ogous to the case of the rod antennas, with the only difference that three modes are considered instead of one. As before, we take a fixed value of $\kappa = (1 + 0.01i) \times 10^{-4}$ (cf. Fig. 5a). Figure 5b (top panel) displays the total ΔCD signal. One can again observe an excellent agreement between the prediction of the modal theory (line) and exact full-wave calculations (dots). Two distinct features can be identified: A peak at the resonance energy of mode 1, and a zero crossing surrounded by large absolute values at the resonance energy of mode 2.

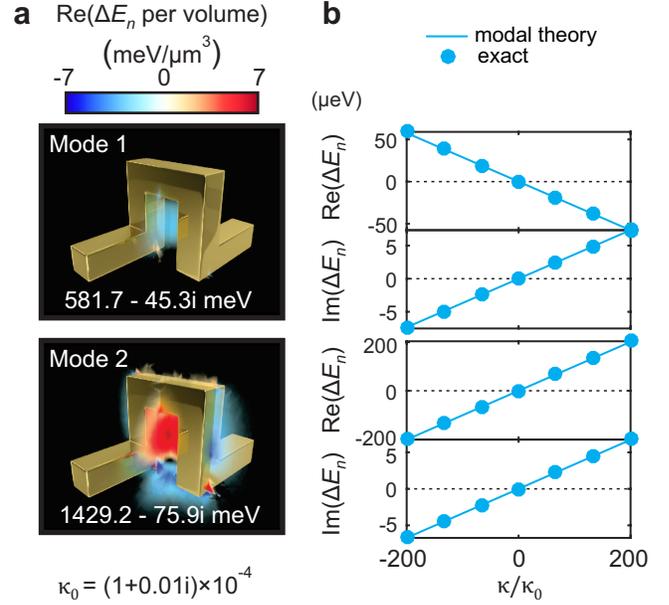

**Fig. 4. Resonance shift in Ω antennas.** The Ω antennas are expected to have a collinear electric and magnetic field at their centers, which results in a nonzero resonance shift. **a** "Shift per volume" of two plasmonic modes. Both modes exhibit hotspots with high values and a uniform sign in the center of the antenna. **b** Change in the energy eigenvalues (lines: modal theory; dots: exact full-wave calculations) of the modes as a function of $\kappa$ for a chiral medium located at the center of the antenna as shown in Fig. 1. The top and bottom plots of each subpanel correspond to the change in the real and imaginary part, respectively.

The lower panels depict the individual contributions, subdivided into their modal origin. As it can be seen, in this system, the shift contribution plays an important role for the total signal. However, in addition, also the changes in the excitation and emission efficiencies are quite strong. One can quantify the importance of the individual contributions for each mode separately: For mode 1, the change in the emission efficiencies is dominating over the shift. This combination leads to the peak shape in the ΔCD spectrum. For mode 2, the shift is dominating over the efficiency changes, leading to the zero-crossing behavior. The contributions associated with mode 3 only play a minor role. The same applies for the nonresonant interaction. In summary, the Ω example demonstrates that it is indeed possible to design a sensor with large resonance shifts. However, interestingly, even in this sensor, the relevance of the changes in the excitation and emission efficiencies should not be underestimated. We do not see any reason that this should be different for other structures optimized for strong chiral near-fields. Thus, while an analysis of optical chirality can yield promising nanostructure designs for sensing applications, all contributions must be taken into account for sensor optimization

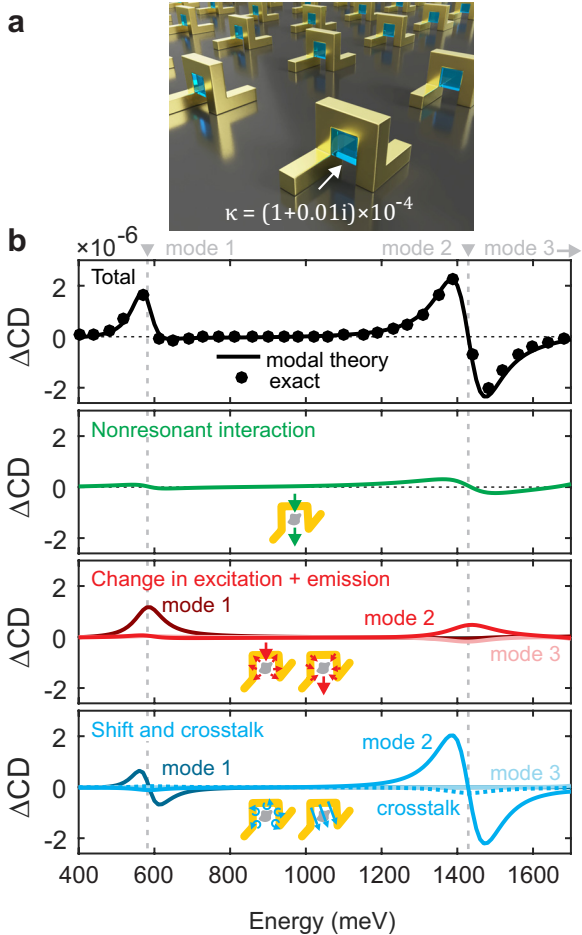

**Fig. 5. Optical response of the Ω antennas. a** An array of antennas with chiral media patches in their center. **b** Resulting ΔCD spectrum. The top plot denotes the total signal (line: calculated with our modal theory; dots: exact full-wave calculations). The individual contributions are depicted below. As it can be seen, in this system, the change in the excitation and emission efficiencies as well as the shift are all relevant. Around mode 2, the resonance shift is dominant.

# Discussion

After considering these two example systems, one might wonder which one has the better overall performance. Therefore, we evaluated their CD enhancement factors[70] (defined as |ΔCD| of the system normalized to the |CD| of the chiral patches without the antennas). The results are depicted in the Supplementary Fig. 7. The answer might appear rather surprising: The rod antennas exhibit a maximum value of 325 and thereby outperform the Ω antennas, which only provide a maximum value of 250. Another advantage of the rod antennas is that, since they are geometrically achiral, they do not provide any CD signal in the absence of the chiral medium[34].

In accordance with a previous numerical study[70], our results suggest that changes in the excitation and emission efficiencies (related to strong overlap of the incident fields with the modes) have to be considered as a relevant mechanism for nanophotonic chiral sensing rather than resonance shifts (related to strong overlap between the modes' electric and magnetic fields). Furthermore, our derivations reveal that in general, stronger near-fields – regardless of with or without high optical chirality – lead to larger signals. There is no need for designing systems such that they exhibit a strong optical chirality with simultaneously weak electric fields, as it was suggested in early works[35,104]. The difference is that these early works focused on optimizing a quantity known as enantioselectivity and not the absolute signal strength, which, however, denotes the relevant factor that defines the detection limits in nanophotonic sensing[16].

All considerations made in this work are based on a first-order approximation in $\kappa$. Therefore, it is quite natural to ask what the limitations of this approach are. To systematically investigate the validity range of the theory, we have varied $\kappa$ over many orders of magnitude and compared the predicted ΔCD spectra to exact full-wave calculations. Negative signs were considered as well. The results are depicted in Supporting Fig. 8. They reveal that the first-order approximation is accurate over a surprisingly huge range of values. Only when the order of $|\kappa|$ approaches unity, the deviations become relevant. Such values would be, however, far beyond what is known for any natural material.

Calculating the spectra for different values of $\kappa$ reveals an additional benefit of the modal theory: While conventional full-wave simulations have to be repeated for multiple $\kappa$ values, the modal theory allows to predict the output over the whole range of values with one single calculation. This is possible, because $\kappa$ appears as a linear factor in the integrals of $\delta S$ [see Eqs. (6) to (11)]. Therefore, one only has to evaluate the integrals with $\kappa$ factored out and can multiply the result with any complex value of interest, to directly obtain the desired spectra. This even works when $\kappa$ is not a constant, but a function of frequency instead. There are at least two applications in sensor modeling: First, one is often interested in the sensor response to different analyte media. Second, even for one particular chiral analyte, one is typically interested in the response to both of its enantiomers (i.e., to both values ±$\kappa$). Related to sensor modeling, there is a further benefit of the perturbative approach: It should not be forgotten that realistic $\kappa$ values are typically extremely small. In full-wave calculations, this sets high standards for the accuracy of the simulations, so that the relevant signals do not vanish within numerical noise[51,70]. The smaller the value of $\kappa$ gets, the more computationally expensive the numerical simulation becomes. In sharp contrast, the perturbative approach can effortlessly predict spectral changes for *arbitrarily* small values of $\kappa$.

In conclusion, we have presented a general theory of chiral light-matter interactions in nanophotonic resonators. Our theory reveals the mechanisms behind nanophotonic chiral sensing. There are exactly five contributions: a nonresonant interaction, changes in the excitation and emission efficiencies of the resonator modes, modal resonance shifts, and intermodal crosstalk. We have investigated the impact of these contributions in different sensor geometries. We have demonstrated that – contrary to common expectation – resonance shifts are often not the dominating source of signal. In the case of achiral sensors, they are even strictly zero. Instead, it turns out that the changes in the excitation and emission efficiencies can be the driving mechanism for enhancing circular dichroism spectroscopy. Besides enabling deep intuitive insights for the understanding and tailoring of nanophotonic chiral light-matter interactions, our theory also constitutes a highly efficient computational tool, with clear advantages over conventional approaches in terms of calculation time and efforts.

# Methods

**Calculations.** All calculations were performed using the commercial finite-element solver COMSOL Multiphysics. The chiral constitutive equations were implemented according to Ref. [70]. The modes were calculated and normalized following the method provided in Ref. [83]. The gold dielectric function was described by a Drude model with plasma frequency $\omega_p = 1.37 \times 10^{16}$ rad/s and a damping constant $\gamma = 1.22 \times 10^{14}$ rad/s (adopted from Ref. [105]). The water was accounted for with its refractive index of 1.33. Calculations have been cross-checked by an in-house implementation of the Fourier-modal method[106].

**Antenna dimensions.** Rod antennas: The rods have a length of 200 nm, a width of 40 nm, a height of 40 nm, and are periodically arranged with a period of 600 nm. The sizes of the chiral media patches are $40 \times 40 \times 40$ nm$^3$. $\Omega$ antennas: The $\Omega$s have a total length of 240 nm, a total width of 140 nm, and a total height of 140 nm. They are wound of a quadratic wire with a lateral extension of 40 nm. The chiral media patches in their centers are 40 nm in length, 60 nm in width, and 60 nm in height. The period is 500 nm.

**Unperturbed scattering matrix.** To improve the accuracy of the unperturbed scattering matrix $S$, the influence of higher-order modes in Eq. (4) was accounted for with a cubic fit, following the method provided in Ref. [80]. The fit was evaluated at four energy points, equidistantly distributed over the depicted spectral range. Note that $S$ is only needed for the calculation of the $\Delta$CD spectra (see equations in the Supporting information), while for predicting the change $\delta S$, which constitutes the main result of this work, it is not required at all.

**Visualization.** The fields and the "shift per volume" densities were displayed on selectively chosen slices through the antennas. The slice plots were generated from simulation data and then incorporated into a three-dimensional model of the structure, which had been created with the open-source graphics suite Blender. The transparency (alpha channel) of each slice plot is proportional to the magnitude of the displayed value. The slice positions were selected such that all relevant features are visible: For the field plot of the rod antenna, the slices are at half of the antenna's width and height; for the corresponding "shift per volume" plot, they are at one quarter and three quarters of the antenna's width and height; for the plots of the $\Omega$ antenna, they are at half the $\Omega$'s width and length.

# Supplementary Information
# Nanophotonic chiral sensing: How does it actually work?

Both et al.

# Supplementary Figures

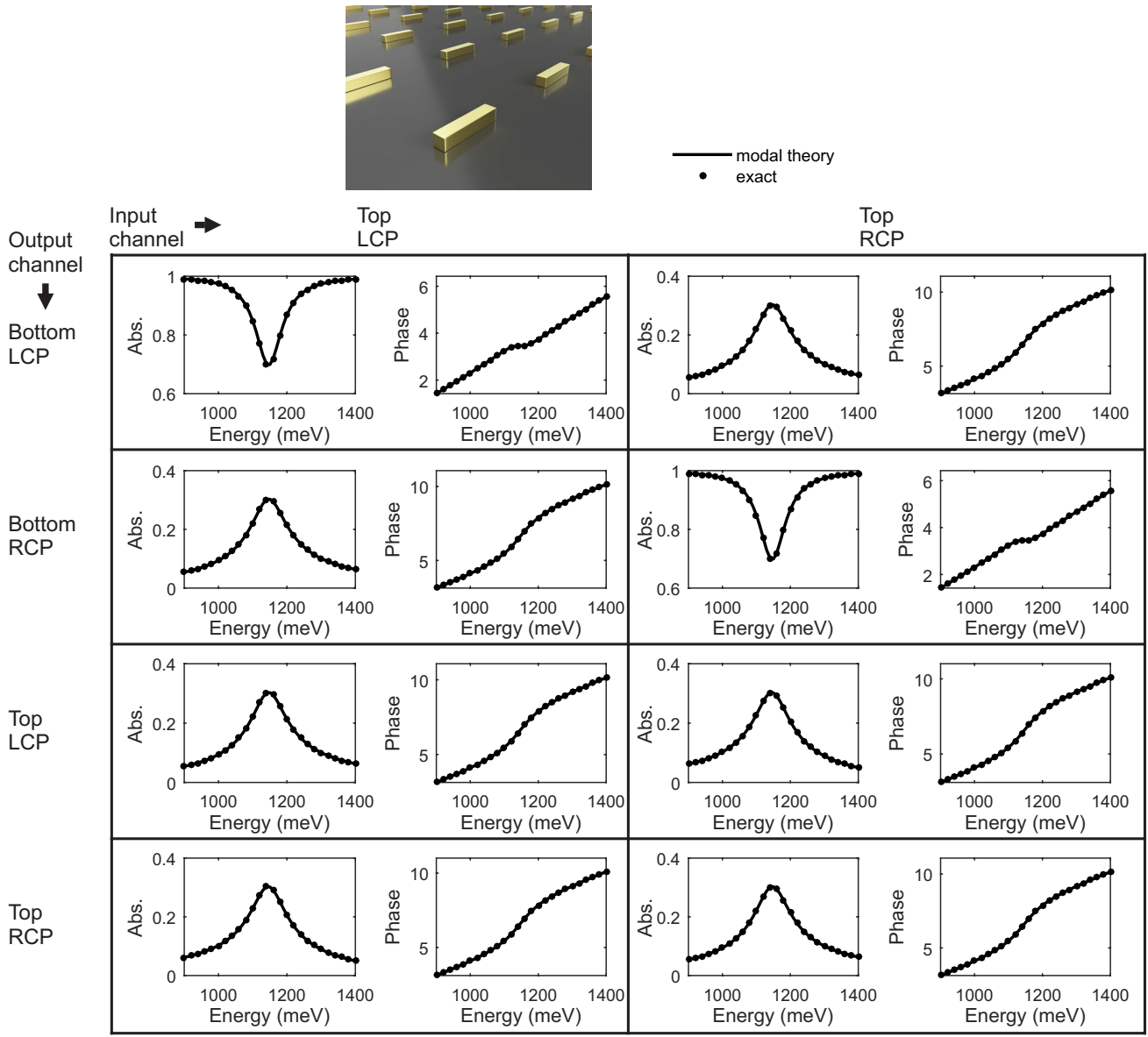

**Supplementary Figure 1. Unperturbed scattering matrix $S$ of the rod antennas.** Each element is represented by its absolute value and its phase. We consider input channels from the top direction and output channels in the top and the bottom direction. The channels are distinguished as left-handed circularly polarized (LCP) and right-handed circularly polarized (RCP). For comparison, we plot both the results of the modal theory and the results of exact full-wave callculations.

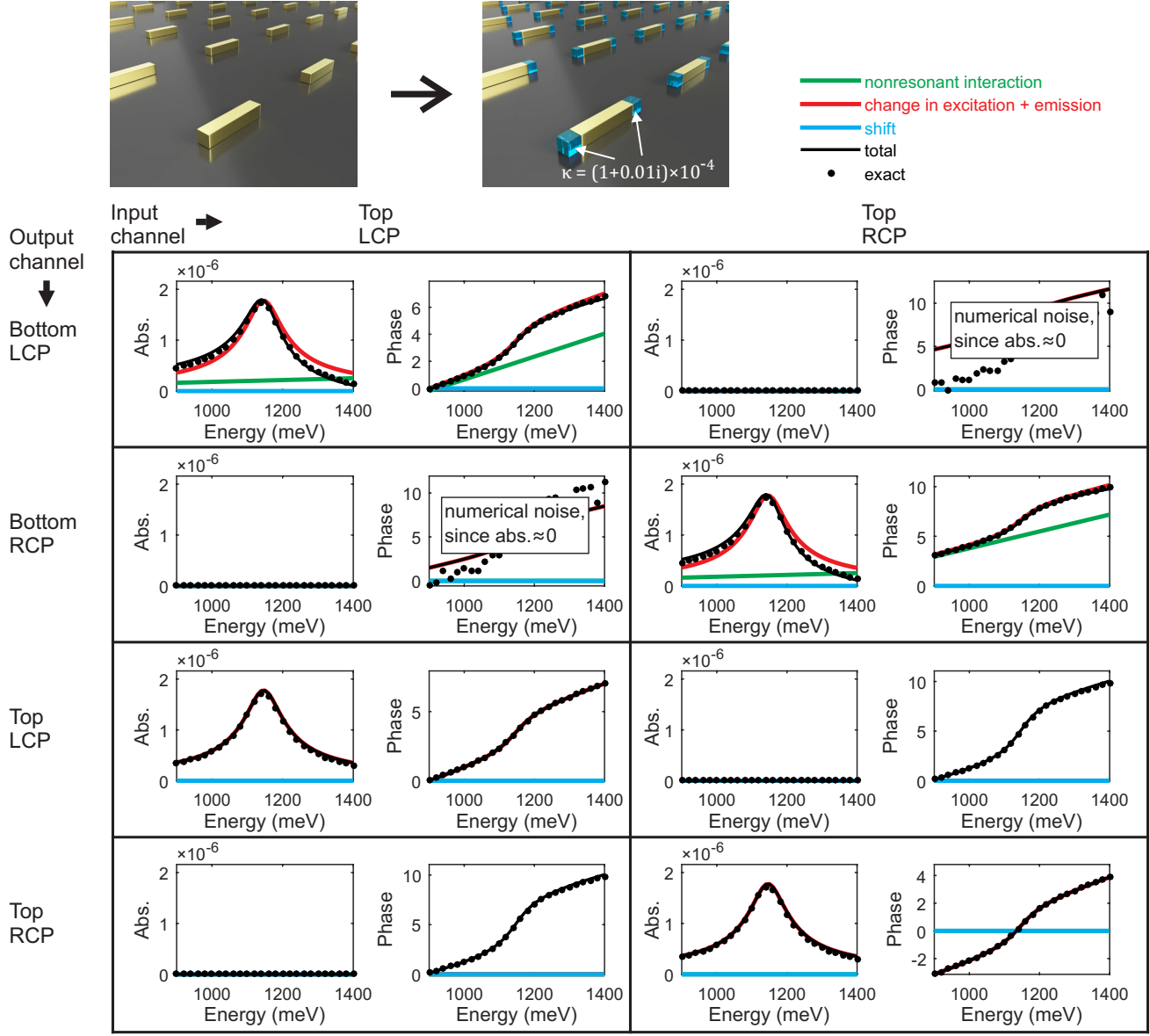

**Supplementary Figure 2. Change of the scattering matrix $\delta S$ of the rod antennas with chiral media patches at each end.** As in Fig. 2b (main manuscript), the lines depict the results of our modal theory (total signal, as well as the separation into individual contributions), while the dots have been obtained from exact full-wave calculations. Note that the (1,2) and the (2,1) matrix components have an absolute value close to zero and hence their phase term is governed by numerical noise. As it can be seen, the shift contribution is zero in all components.

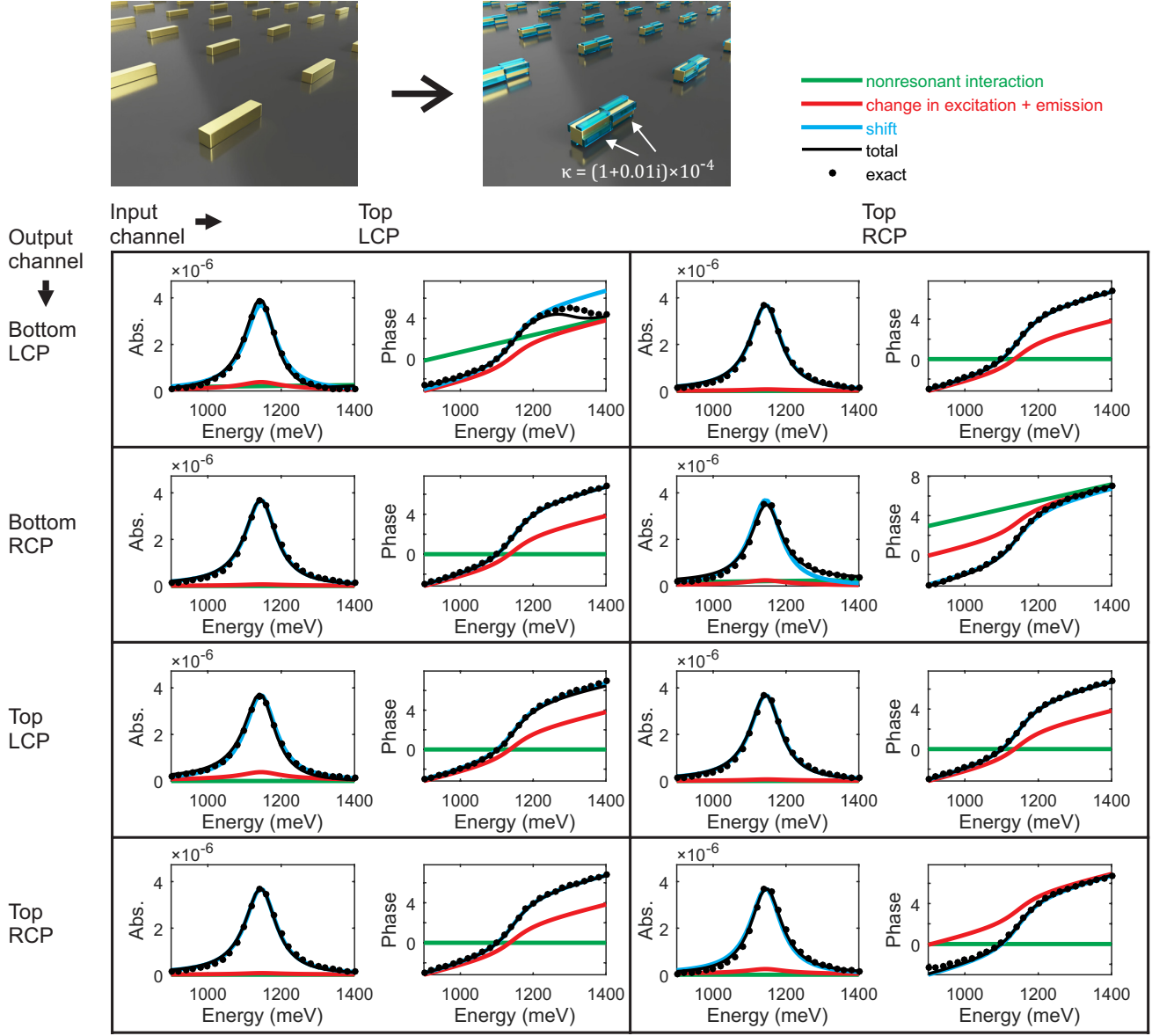

**Supplementary Figure 3. Change of the scattering matrix $\delta S$ of the rod antennas with the chiral medium distributed over chiral hotspots**. In order to enforce a resonance shift in the rod antennas, we take the chiral medium from the end of the antennas [cf. Fig 3b (main manuscript) and Supplementary Fig. 2] and redistribute it over the regions with positive "shift per volume" values [red spots in Fig. 3b (main manuscript)]. As it can be seen, this configuration results in a nonzero shift contribution in all matrix elements.
4...

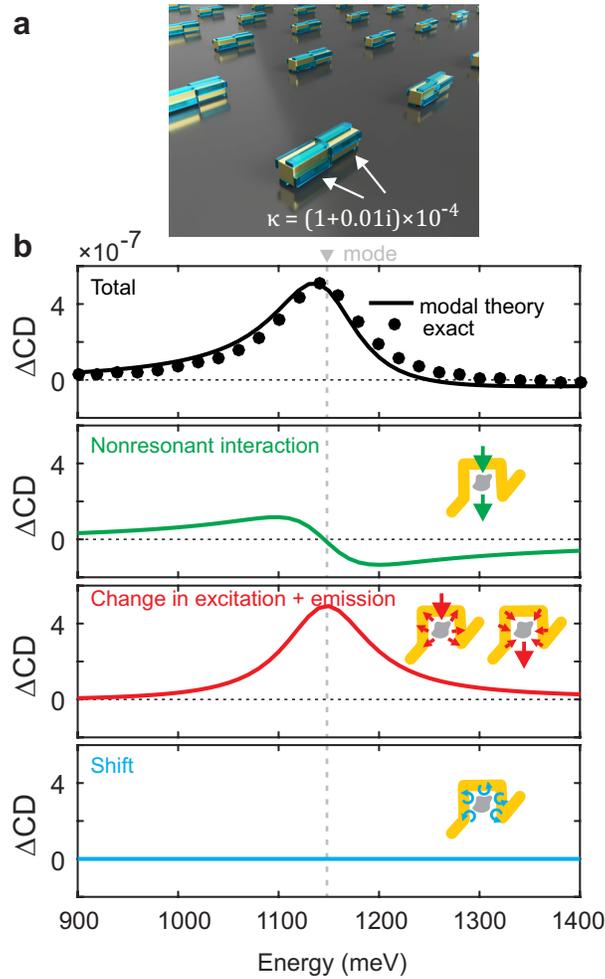

**Supplementary Figure 4. Optical ΔCD response of the rod antennas with the chiral medium distributed over chiral hotspots.** Remarkably, although all matrix elements in Supplementary Fig. 3 exhibit a nonzero shift contribution, the shift contribution in the ΔCD spectrum remains zero. At first glance this might seem surprising; however, it can be understood by noting that the shift contributions in the matrix are symmetric such that they affect left-handed and right-handed circularly polarized input the same way and hence cancel out each other in the ΔCD signal.

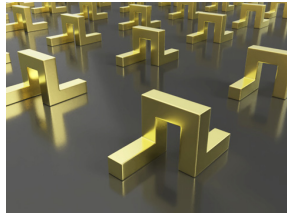
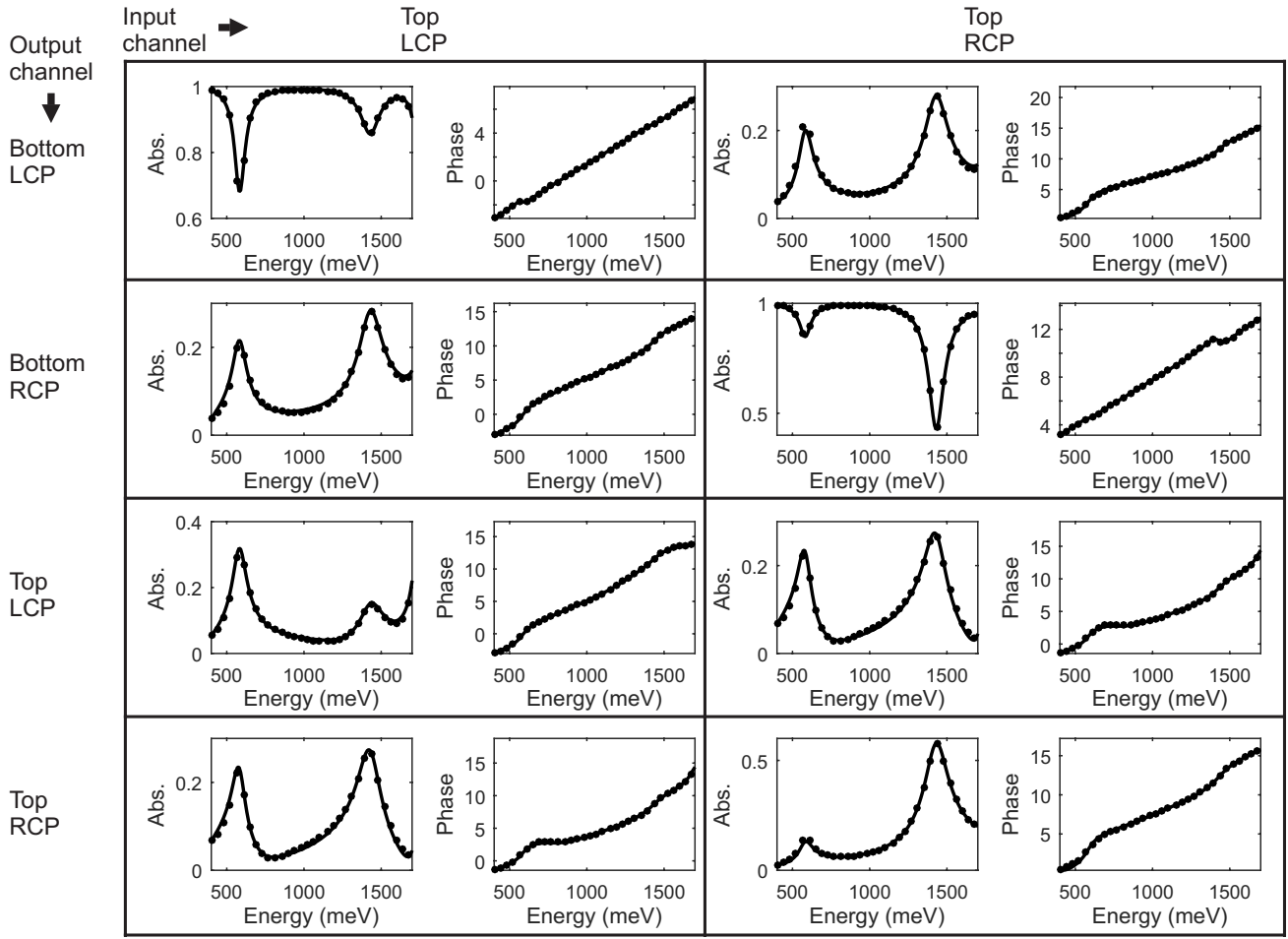

**Supplementary Figure 5. Unperturbed scattering matrix $S$ of the $\Omega$ antennas.**

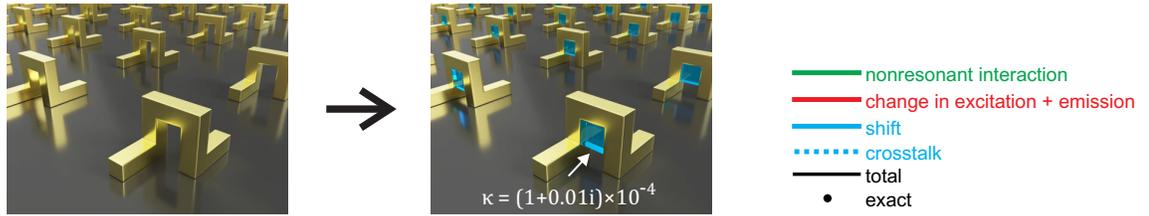
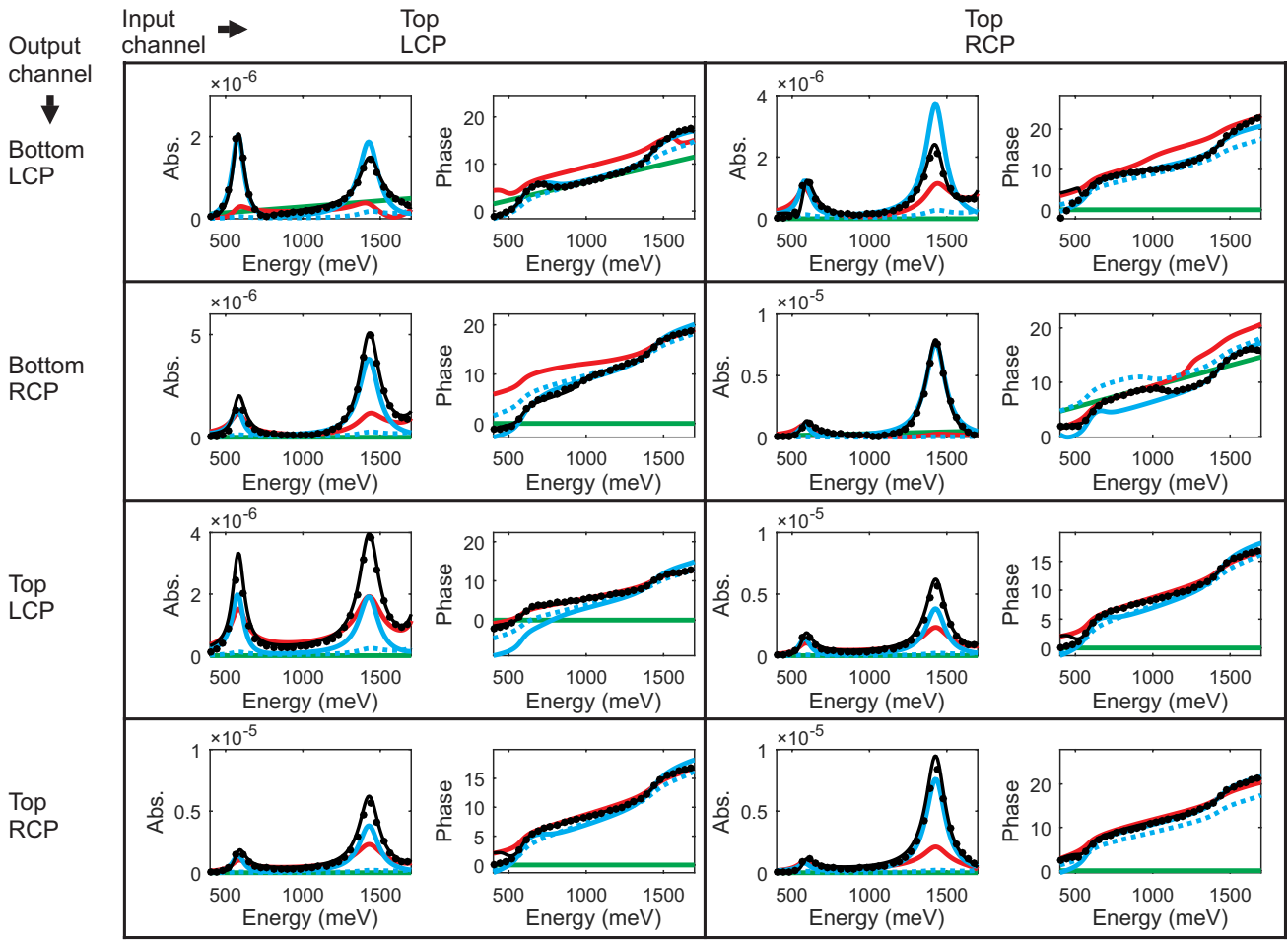

**Supplementary Figure 6. Change of the scattering matrix $\delta S$ of the $\Omega$ antennas with chiral media patches in their centers.**

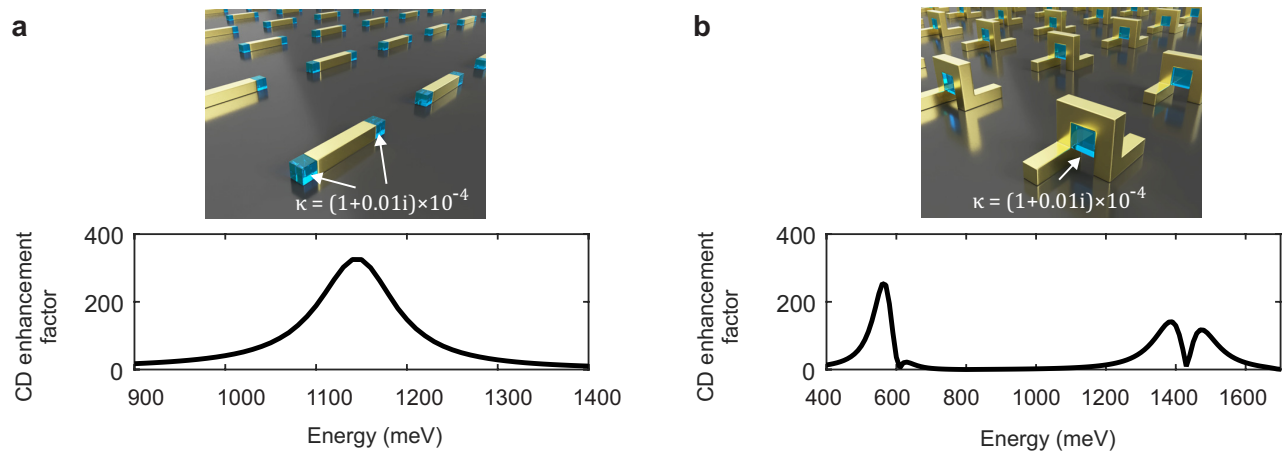

**Supplementary Figure 7. Calculated CD enhancement factors. a** Rod antennas. **b** Ω antennas. The CD enhancement factor was calculated following Ref. [1] as the ratio between the |ΔCD| spectrum of the chiral patches with antennas and the |CD| spectrum of the chiral patches alone.

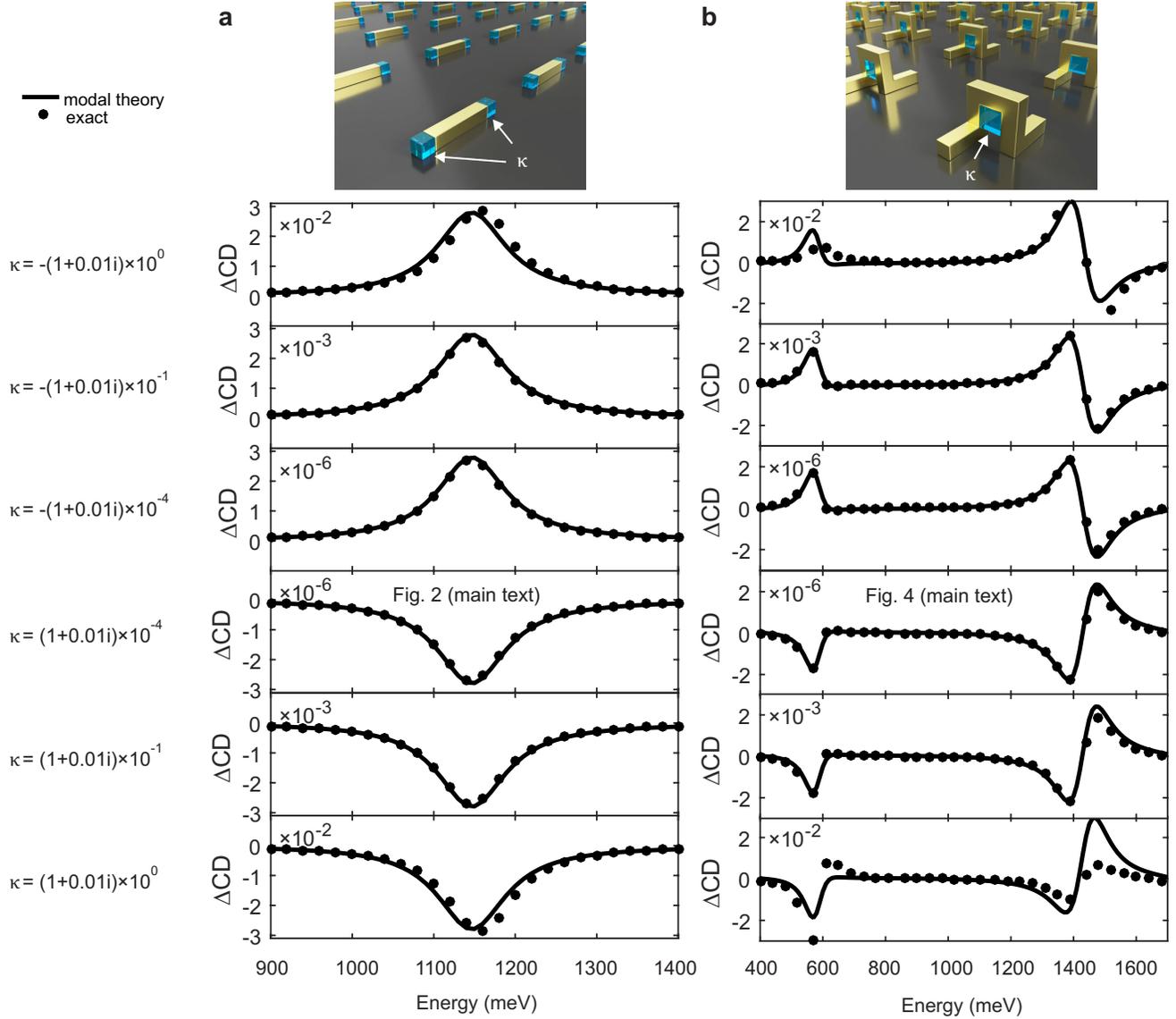

**Supplementary Figure 8. Validity range of the first-order approximation. a** Rod antennas. **b** $\Omega$ antennas. The Pasteur parameter $\kappa$ was varied over many orders of magnitude. Negative signs were considered as well. As it can be seen, our approach works well over a huge range of values. Only when the order of $|\kappa|$ starts to approach unity, the deviations become clearly visible. However, such values would be far beyond what is known for any natural material.

# Supplementary Notes

## 1 Details on the modal theory

In this section, we provide the details on the modal theory. First, we recap from literature[2,3], how the scattering matrix of an open optical resonator can be calculated from its modes. Afterwards, we derive an expression for the change of the scattering matrix under material perturbations of the resonator.

### 1.1 General definitions

We start with curl Maxwell's equations of the unperturbed resonator. By making use of the compact operator formulation introduced in Ref. [4], they can be written as [Gaussian units, frequency domain, time dependence $\exp(-i\omega t)$]

$$\left[ k \underbrace{\begin{pmatrix} \varepsilon(\mathbf{r};k) & -i\xi(\mathbf{r};k) \\ i\zeta(\mathbf{r};k) & \mu(\mathbf{r};k) \end{pmatrix}}_{\hat{\mathbb{P}}(\mathbf{r};k)} - \underbrace{\begin{pmatrix} 0 & \nabla\times \\ \nabla\times & 0 \end{pmatrix}}_{\hat{\mathbb{D}}(\mathbf{r})} \right] \underbrace{\begin{pmatrix} \mathbf{E}(\mathbf{r};k) \\ i\mathbf{H}(\mathbf{r};k) \end{pmatrix}}_{\mathbb{F}(\mathbf{r};k)} = \underbrace{\begin{pmatrix} \mathbf{J}_\mathrm{E}(\mathbf{r};k) \\ i\mathbf{J}_\mathrm{H}(\mathbf{r};k) \end{pmatrix}}_{\mathbb{J}(\mathbf{r};k)}. \tag{1}$$

$$\underbrace{\phantom{x}}_{\hat{\mathbb{M}}(\mathbf{r};k)}$$

Here, $k = \omega/c$ denotes the wavenumber, while $\varepsilon$, $\mu$, $\xi$ and $\zeta$ represent the material parameters, namely permittivity, permeability, and possible bi-anisotropic contributions, respectively. The material parameters can in general be tensors. Note that $\xi$ and $\zeta$ are assumed to be zero in Eq. (2) of the main manuscript, since they are zero in most cases. Furthermore, we assume that the materials are reciprocal[4], leading to $\varepsilon^\mathrm{T} = \varepsilon$, $\mu^\mathrm{T} = \mu$, and $\xi^\mathrm{T} = -\zeta$, where the superscript T denotes the matrix transpose. The vectors $\mathbf{E}$ and $\mathbf{H}$ represent the electric and magnetic fields, respectively, while the vectors $\mathbf{J}_\mathrm{E}$ and $\mathbf{J}_\mathrm{H}$ denote external electric and magnetic currents, respectively. All material parameters are included in an operator $\hat{\mathbb{P}}$. In analogy, all curls are included in an operator $\hat{\mathbb{D}}$. Furthermore, we introduced two six-dimensional supervectors $\mathbb{F}$ and $\mathbb{J}$ that consist of the fields and the currents, respectively. By further abbreviating $\hat{\mathbb{M}} = k\hat{\mathbb{P}} - \hat{\mathbb{D}}$, the curl Maxwell's equations become a simple operator equation

$$\hat{\mathbb{M}}(\mathbf{r};k)\mathbb{F}(\mathbf{r};k) = \mathbb{J}(\mathbf{r};k). \tag{2}$$

For later convenience, we follow Ref. [2] and define two types of bilinear maps between two arbitrary field supervectors $\mathbb{A}$ and $\mathbb{B}$ with

$$\mathbb{A} = \begin{pmatrix} \mathbf{E}_\mathrm{A} \\ i\mathbf{H}_\mathrm{A} \end{pmatrix}, \qquad\qquad \mathbb{B} = \begin{pmatrix} \mathbf{E}_\mathrm{B} \\ i\mathbf{H}_\mathrm{B} \end{pmatrix}. \tag{3}$$

The first bilinear map represents a volume integral over a finite volume $V$:

$$\langle \mathbb{A}|\mathbb{B} \rangle_V \equiv \int_V dV \left( \mathbf{E}_\mathrm{A} \cdot \mathbf{E}_\mathrm{B} - \mathbf{H}_\mathrm{A} \cdot \mathbf{H}_\mathrm{B} \right). \tag{4}$$

The second one represents a surface integral over the surrounding surface $\partial V$ of $V$:

$$[\mathbb{A}|\mathbb{B}]_{\partial V} \equiv i \oint_{\partial V} d\mathbf{S} \cdot \left( \mathbf{E}_\mathrm{A} \times \mathbf{H}_\mathrm{B} - \mathbf{E}_\mathrm{B} \times \mathbf{H}_\mathrm{A} \right). \tag{5}$$

It can be shown that[2]

$$[\mathbb{A}|\mathbb{B}]_{\partial V} = \langle \mathbb{B}|\hat{\mathbb{D}}|\mathbb{A} \rangle_V - \langle \mathbb{A}|\hat{\mathbb{D}}|\mathbb{B} \rangle_V, \tag{6}$$

which will be useful later.

### 1.2 Green's dyadic and modes

Now, let us come back to Eq. (2). One can introduce the so-called Green's dyadic $\hat{\mathbb{G}}$ of this equation, which is defined as the solution of[4]

$$\hat{\mathbb{M}}(\mathbf{r};k)\hat{\mathbb{G}}(\mathbf{r},\mathbf{r}';k) = \mathbb{1}\delta(\mathbf{r}-\mathbf{r}'), \tag{7}$$

where $\mathbb{1}$ denotes the $6 \times 6$ unit matrix and $\delta(\mathbf{r})$ is the delta function. With the help of $\hat{\mathbb{G}}$, one can formally solve Eq. (2) for the fields $\mathbb{F}$ induced by a given source source $\mathbb{J}$ via[2,4]

$$\mathbb{F}(\mathbf{r};k) = \int_V dV' \hat{\mathbb{G}}(\mathbf{r},\mathbf{r}';k) \mathbb{J}(\mathbf{r}';k). \tag{8}$$

The Green's dyadic $\hat{\mathbb{G}}$ can be expanded in terms of the modes. These are solution of Eq. (2) for $\mathbb{J} = 0$ and outgoing boundary conditions:

$$\hat{\mathbb{M}}(\mathbf{r},k_n) \mathbb{F}_n(\mathbf{r}) = 0. \tag{9}$$

Here, $\mathbb{F}_n$ represent the modes and their corresponding wavenumber $k_n$ eigenvalues. Note that Eq. (9) defines $\mathbb{F}_n$ only up to an arbitrary scalar factor. In order to be used in an expansion, the resonant field distributions have to be normalized. Valid normalization schemes can be found in Refs. [2, 4, 5] and references therein. With the normalized modes, the Green's dyadic can be written as[2,4]

$$\hat{\mathbb{G}}(\mathbf{r},\mathbf{r}';k) = \sum_n \frac{\mathbb{F}_n(\mathbf{r}) \otimes \mathbb{F}_n^R(\mathbf{r}')}{k - k_n}. \tag{10}$$

Here, the superscript R denotes the reciprocal conjugate modes[2]. Note that, although the term may suggest otherwise, this has nothing to do with the reciprocity of the medium. Reciprocal conjugate fields are solutions of the same Maxwell's equations at identical frequencies that have to be determined in dependence of the geometry in order to warrant the validity of the resonant expansion in the case of degeneracy. For spherically symmetric systems, reciprocal conjugation means switching the azimuthal order from $m$ to $-m$. For planar periodic systems, this means switching the in-plane wave vector from $\mathbf{k}_\parallel$ to $-\mathbf{k}_\parallel$. In the case of planar periodic systems with normal incidence (this applies for the example systems from the main manuscript), one finds $\mathbf{k}_\parallel = 0$, which trivially results in $\mathbb{F}_n^R = \mathbb{F}_n$.

Note that Eq. (10) neglects so-called cut contributions[6–8], which represent a continuum of states that appears, e.g., in two-dimensional systems. Furthermore, note that the expansion of $\hat{\mathbb{G}}$ in terms of the modes is strictly speaking only complete inside the resonant structure, but not in the distant surrounding[9]. In numerical calculations, it is possible to account for the cut contributions, as well as to ensure completenes over the whole space by terminating the calculation domain with so-called perfectly matched layers (PMLs) and include so-called PML modes in the expansion[5,10]. PML modes are included the same way as other modes.

### 1.3 Scattered field formalism

Let us now introduce the so-called scattered field formalism[2,11–14]. In general, a nanophotonic resonator consist of a resonating entity – from now on referred to as scatterer – and a background system, in which this scatter is embedded. Typical examples for scatterers are nanoparticles, nanoantennas, arrays of nanoantennas, or gratings. Typical examples for background systems are vacuum, homogeneous media, or substrate/superstrate interfaces. Consequently, one can split the material distribution $\hat{\mathbb{P}}$ of the resonator into a part $\hat{\mathbb{P}}_{bg}$ that describes the background system and a part $\Delta\hat{\mathbb{P}} = \hat{\mathbb{P}} - \hat{\mathbb{P}}_{bg}$ that describes the scatterer. Note that $\hat{\mathbb{P}}_{bg}$ is nonzero everywhere in space, while $\Delta\hat{\mathbb{P}}$ is nonzero only inside the scatterer. We then consider Maxwell's equations of the resonator without external currents (i.e., $\hat{\mathbb{M}}\mathbb{F} = 0$). The total field $\mathbb{F}$ in the resonator can be separated into a background field and a scattered field: $\mathbb{F} = \mathbb{F}_{bg} + \mathbb{F}_{scat}$. The background field $\mathbb{F}_{bg}$ is defined as a solution of Maxwell's equation in the background system and satisfies

$$\hat{\mathbb{M}}_{bg} \mathbb{F}_{bg} = 0, \tag{11}$$

with $\hat{\mathbb{M}}_{bg} \equiv k\hat{\mathbb{P}}_{bg} - \hat{\mathbb{D}}$. The scattered field $\mathbb{F}_{scat}$ denotes the response of the scatterer to the background field. After some algebra, one finds

$$\hat{\mathbb{M}} \mathbb{F}_{scat} = -\Delta\hat{\mathbb{M}} \mathbb{F}_{bg}, \tag{12}$$

with $\Delta\hat{\mathbb{M}} \equiv k\Delta\hat{\mathbb{P}}$.

Equation (12) is formally identical to Eq. (2) (with $\mathbb{F}_{scat}$ taking the role of $\mathbb{F}$ and $-\Delta\hat{\mathbb{M}}\mathbb{F}_{bg}$ taking the role of a source $\mathbb{J}$). Hence, we can solve Eq. (12) for $\mathbb{F}_{scat}$ with the help of Eq. (8) using the Green's dyadic provided in Eq. (10). This result in an expansion of the scattered field[2]

$$\mathbb{F}_{scat} = -\sum_n \mathbb{F}_n \frac{\langle \mathbb{F}_n^R | \Delta\hat{\mathbb{M}} | \mathbb{F}_{bg} \rangle_V}{k - k_n}, \tag{13}$$

which will be used later.

## 1.4 Scattering matrix

We now choose a surface $\partial V$ that surrounds our scatterer. On this surface, one can construct a complete and orthogonal set of incoming $\{\mathbb{I}_\mathbf{N}\}$ and outgoing $\{\mathbb{O}_\mathbf{N}\}$ basis functions of the background system that allows to decompose an arbitrary field[2,15]. Here, the vector $\mathbf{N}$ denotes a set of quantum numbers that is used to label the basis functions. They satisfy the following orthogonality relations[2]:

$$[\mathbb{I}_\mathbf{N}^R|\mathbb{O}_{\mathbf{N}'}]_{\partial V} = -[\mathbb{O}_\mathbf{N}^R|\mathbb{I}_{\mathbf{N}'}]_{\partial V} = \delta_{\mathbf{N},\mathbf{N}'},$$
$$[\mathbb{I}_\mathbf{N}^R|\mathbb{I}_{\mathbf{N}'}]_{\partial V} = [\mathbb{O}_\mathbf{N}^R|\mathbb{O}_{\mathbf{N}'}]_{\partial V} = 0, \quad (14)$$

where the superscript R labels the reciprocal conjugate basis functions[2]. For single-particle scatteres, it is convenient to take $\partial V$ as a sphere around the scatterer. In this case, the appropriate basis functions are vector spherical harmonics. For planar periodic scatterers, it is convenient, to take $\partial V$ as two planes, one above and one below the scatterer. In this case, the appropriate basis functions are plane waves. Details can be found in Ref. [2]. Incoming basis functions $\{\mathbb{I}_\mathbf{N}\}$ carry energy towards the scatterer, while outgoing basis functions $\{\mathbb{O}_\mathbf{N}\}$ carry energy away from the scatterer. For every incoming basis function $\mathbb{I}_\mathbf{N}$ there exists an outgoing counterpart $\mathbb{O}_\mathbf{N}$. Each incoming basis function defines a so-called incoming channel, while each outgoing basis function defines a so-called outgoing channel. For a given resonator, the relationship between incoming and outgoing channels can be summarized via the so-called scattering matrix $S$. Each element $S_\mathbf{MN}$ of the scattering matrix represents the transmission (or reflection) amplitude from one particular input channel $\mathbf{N}$ into one particular output channel $\mathbf{M}$.

Let us now show how $S_\mathbf{MN}$ can be calculated from the modes. Therefore, we assume that the resonator is excited by an incoming channel $\mathbf{N}$. This excitation generates a background field and scattered field, which we label as $\mathbb{F}_{\text{bg},\mathbf{N}}$ and $\mathbb{F}_{\text{scat},\mathbf{N}}$, respectively. The fields $\mathbb{B}$ and $\mathbb{F}_{\text{scat},\mathbf{N}}$ then excite outgoing channels of the resonator. The coupling into a particular output channel $\mathbf{M}$ determines the element $S_\mathbf{MN}$ of the scattering matrix. It can be calculated by projecting the fields onto the probe function $\mathbb{I}_\mathbf{M}^R$ (reciprocal conjugate of the incoming version of $\mathbb{O}_\mathbf{M}$) via[2]

$$S_\mathbf{MN} = [\mathbb{I}_\mathbf{M}^R|\mathbb{F}_{\text{bg},\mathbf{N}}]_{\partial V} + [\mathbb{I}_\mathbf{M}^R|\mathbb{F}_{\text{scat},\mathbf{N}}]_{\partial V}. \quad (15)$$

The first term on the right-hand side describes the direct coupling between incoming and outgoing channels in the background system, while the second term describes the interaction between incoming and outgoing channels induced by the scatterer. In the following, we will assume that the background field $\mathbb{F}_{\text{bg},\mathbf{N}}$ is known and only focus on the scattered field $\mathbb{F}_{\text{scat},\mathbf{N}}$. Details on how to obtain $\mathbb{F}_{\text{bg},\mathbf{N}}$ for an input channel $\mathbb{I}_\mathbf{N}$ in a given resonator can be found, e.g., in Ref. [3].

Let us now derive a more explicit expression for the second term on the right-hand side. Let $\mathbb{F}_{\text{bg},\mathbf{M}}^R$ be the resulting background field that would be obtained by launching the probe function $\mathbb{I}_\mathbf{M}^R$ into the background system. Then we can convert:

$$[\mathbb{I}_\mathbf{M}^R|\mathbb{F}_{\text{scat},\mathbf{N}}]_{\partial V} \stackrel{(*)}{=} [\mathbb{F}_{\text{bg},\mathbf{M}}^R|\mathbb{F}_{\text{scat},\mathbf{N}}]_{\partial V} \stackrel{(**)}{=} -\langle\mathbb{F}_{\text{bg},\mathbf{M}}^R|\Delta\hat{\mathbb{M}}|\mathbb{F}_{\text{bg},\mathbf{N}}\rangle_V - \langle\mathbb{F}_{\text{bg},\mathbf{M}}^R|\Delta\hat{\mathbb{M}}|\mathbb{F}_{\text{scat},\mathbf{N}}\rangle_V, \quad (16)$$

where $V$ denotes the volume that is surrounded by $\partial V$. For the first step (*), we used that on the surface $\partial V$, the scattered field $\mathbb{F}_{\text{scat},\mathbf{N}}$ is purely composed of outgoing basis functions[2,3], while $\mathbb{F}_{\text{bg},\mathbf{M}}^R$ contains one incoming basis function (namely $\mathbb{I}_\mathbf{M}^R$) plus some superposition of outgoing basis functions[2,3], and applied the the orthogonality relations from Eq. (14). For the second step (**), we used Eq. (6) to convert the surface integral into volume integrals; then, we inserted $\hat{\mathbb{D}} = k\hat{\mathbb{P}} - \hat{\mathbb{M}}$ and $\hat{\mathbb{M}} = \hat{\mathbb{M}}_{\text{bg}} + \Delta\hat{\mathbb{M}}$, resulting in $\langle\mathbb{F}_{\text{scat},\mathbf{N}}|k\hat{\mathbb{P}} - \hat{\mathbb{M}}_{\text{bg}} - \Delta\hat{\mathbb{M}}|\mathbb{F}_{\text{bg},\mathbf{M}}^R\rangle_V - \langle\mathbb{F}_{\text{bg},\mathbf{M}}^R|k\hat{\mathbb{P}} - \hat{\mathbb{M}}|\mathbb{F}_{\text{scat},\mathbf{N}}\rangle_V$; next, we applied Eqs. (11) and (12) and furthermore used that the materials are reciprocal, which allows to switch the under the integrals as $\langle\mathbb{F}_{\text{scat},\mathbf{N}}|k\hat{\mathbb{P}}|\mathbb{F}_{\text{bg},\mathbf{M}}^R\rangle_V = \langle\mathbb{F}_{\text{bg},\mathbf{M}}^R|k\hat{\mathbb{P}}|\mathbb{F}_{\text{scat},\mathbf{N}}\rangle_V$ and $\langle\mathbb{F}_{\text{scat},\mathbf{N}}|\Delta\hat{\mathbb{M}}|\mathbb{F}_{\text{bg},\mathbf{M}}^R\rangle_V = \langle\mathbb{F}_{\text{bg},\mathbf{M}}^R|\Delta\hat{\mathbb{M}}|\mathbb{F}_{\text{scat},\mathbf{N}}\rangle_V$.

Plugging Eq. (16) into Eq. (15) and inserting the expansion of the scattered field [Eq. (13)] results in Eq. (4) of the main manuscript:

$$S_\mathbf{MN} = \underbrace{[\mathbb{I}_\mathbf{M}^R|\mathbb{F}_{\text{bg},\mathbf{N}}]_{\partial V} - \langle\mathbb{F}_{\text{bg},\mathbf{M}}^R|\Delta\hat{\mathbb{M}}|\mathbb{F}_{\text{bg},\mathbf{N}}\rangle_V}_{\equiv S_\mathbf{MN}^{\text{bg}}} + \sum_n \frac{\overbrace{\langle\mathbb{F}_{\text{bg},\mathbf{M}}^R|\Delta\hat{\mathbb{M}}|\mathbb{F}_n\rangle_V}^{\equiv a_{n,\mathbf{M}}} \overbrace{\langle\mathbb{F}_n^R|\Delta\hat{\mathbb{M}}|\mathbb{F}_{\text{bg},\mathbf{N}}\rangle_V}^{\equiv b_{n,\mathbf{N}}}}{k - k_n}. \quad (17)$$

This equation allows to calculate the scattering matrix of an open optical resonator from the knowledge of its modes. Here, $S_\mathbf{MN}^{\text{bg}}$ corresponds to a nonresonant background term, while $a_{n,\mathbf{M}}$ and $b_{n,\mathbf{N}}$ represent the emission and excitation coefficients, respectively, of the modes $\mathbb{F}_n$, and $k$ denotes the wavenumber, at which the resonator is excited. Note that all quantities except $\mathbb{F}_n$ and $k_n$ are $k$ dependent. We want to remark that this equation is similar to Eq. (34) from Ref. [3]. Furthermore, we want to point out that the first part of $S_\mathbf{MN}^{\text{bg}}$ represents the scattering matrix of the background system, while the second part can be associated with a Born-like scattering process[3].

## 1.5 Change of the scattering matrix

In the following, we derive an equation for the change of the scattering matrix under material perturbations of the resonator. Let the operator $\hat{\mathbb{M}}$ as before represent the unperturbed resonator. Now let us assume the resonator gets perturbed by some local material changes. The perturbed resonator is then described by a new operator

$$\hat{\mathbb{M}}_{\text{pert}} = \hat{\mathbb{M}} + \Lambda \delta\hat{\mathbb{M}}, \tag{18}$$

where $\delta\hat{\mathbb{M}}$ is an operator that describes the perturbation and $\Lambda$ is a dimensionless parameter that allows to switch the perturbation on ($\Lambda = 1$) and off ($\Lambda = 0$). Although in the context of this work, we are only interested in perturbations connected to the chirality parameter $\kappa$, we make our derivations more general and allow $\delta\hat{\mathbb{M}}$ to contain any kind of material perturbation. The only assumption we make is that the change is restricted to some finite volume $V_c$ and that $V_c$ is close enough to the scatterer that the expansion in Eq. (13) is still valid. The perturbation then reads as

$$\delta\hat{\mathbb{M}} = \begin{cases} k \begin{pmatrix} \delta\varepsilon & -i\delta\xi \\ i\delta\zeta & \delta\mu \end{pmatrix} & \text{inside volume } V_c, \\ 0 & \text{outside}, \end{cases} \tag{19}$$

where $\delta\varepsilon$, $\delta\mu$, $\delta\xi$ and $\delta\zeta$ denote the changes in the permittivity, the permeability and the bi-anisotropic contributions, respectively. Note that in general these quantities can be tensors. For the special case discussed in the main manuscript, where a resonator with $\kappa = 0$ is changed to a resonator with $\kappa \neq 0$, one gets $\delta\varepsilon = \delta\mu = 0$ and $-i\delta\xi = i\delta\zeta = -\underline{1}\kappa$, where $\underline{1}$ denotes the $3 \times 3$ unit matrix.

Let $\mathbb{F}_{\text{pert}}$ now denote the total field in the perturbed resonator. In analogy to the unperturbed case discussed above, we split $\mathbb{F}_{\text{pert}}$ into a background field and a scattered field. Only the scattered field is affected by the perturbation, while the background field remains unaffected. As in conventional perturbative approaches (e.g., from quantum mechanics), we write $\mathbb{F}_{\text{pert}}$ as a Taylor series in $\Lambda$:

$$\mathbb{F}_{\text{pert}} = \mathbb{F}_{\text{bg}} + \mathbb{F}_{\text{scat}} + \Lambda \mathbb{F}_{\text{scat}}^{(1)} + \Lambda^2 \mathbb{F}_{\text{scat}}^{(2)} + \ldots, \tag{20}$$

where $\mathbb{F}_{\text{bg}}$ and $\mathbb{F}_{\text{scat}}$ denote the background and scattered field from the unperturbed case, respectively, and $\mathbb{F}_{\text{scat}}^{(1)}$, $\mathbb{F}_{\text{scat}}^{(2)}$, ... are correction terms for the scattered field. The total perturbed field $\mathbb{F}_{\text{pert}}$ fulfills $\hat{\mathbb{M}}_{\text{pert}} \mathbb{F}_{\text{pert}} = 0$. Inserting the expressions for $\hat{\mathbb{M}}_{\text{pert}}$ and $\mathbb{F}_{\text{pert}}$ and comparing the coefficients for every power of $\Lambda$ results in:

$$\hat{\mathbb{M}} \mathbb{F}_{\text{scat}} = -\Delta\hat{\mathbb{M}} \mathbb{F}_{\text{bg}} \tag{21}$$

$$\hat{\mathbb{M}} \mathbb{F}_{\text{scat}}^{(1)} = -\delta\hat{\mathbb{M}} \mathbb{F}_{\text{bg}} - \delta\hat{\mathbb{M}} \mathbb{F}_{\text{scat}} \tag{22}$$

$$\vdots$$

This set of equations implicitly describes the scattered field and its correction terms. The first equation defines $\mathbb{F}_{\text{scat}}$ in the unperturbed resonator and is already known from above [see Eq. (12)], where we had solved it with the help of the Green's dyadic to get an explicit expression for $\mathbb{F}_{\text{scat}}$ [see Eq. (13)]. The second equation defines the first-order correction term $\mathbb{F}_{\text{scat}}^{(1)}$. In analogy to above, we solve it with the help of the Green's dyadic. This gives

$$\mathbb{F}_{\text{scat}}^{(1)} = -\sum_n \mathbb{F}_n \frac{\langle \mathbb{F}_n^R | \delta\hat{\mathbb{M}} | \mathbb{F}_{\text{bg}} \rangle_V}{k - k_n} - \sum_n \mathbb{F}_n \frac{\langle \mathbb{F}_n^R | \delta\hat{\mathbb{M}} | \mathbb{F}_{\text{scat}} \rangle_V}{k - k_n}. \tag{23}$$

Successively applying this method allows to derive expressions for all higher-order correction terms $\mathbb{F}_{\text{scat}}^{(2)}$, ...; however, we will now assume that our perturbation $\delta\hat{\mathbb{M}}$ is small compared to the unperturbed material parameters of the resonator. For the case of chiral perturbations, this is a very justified assumption. Hence, it is enough to consider only the first-order correction term, and the perturbed field can be approximated as $\mathbb{F}_{\text{pert}} \approx \mathbb{F}_{\text{bg}} + \mathbb{F}_{\text{scat}} + \mathbb{F}_{\text{scat}}^{(1)}$. Let us now calculate the scattering matrix of the perturbed resonator. As before in the unperturbed case, we assume that the resonator is excited by a channel $\mathbb{I}_{\mathbf{N}}$, and project the resulting fields onto the probe function $\mathbb{I}_{\mathbf{M}}^R$. Note that the integration surface $\partial V$ for the projection has to be chosen large enough to include all inhomogeneities of the resonator (i.e., the regions where $\Delta\hat{\mathbb{M}}$ and $\delta\hat{\mathbb{M}}$ are nonzero). The perturbed scattering matrix is then obtained as

$$S_{\mathbf{MN}}^{\text{pert}} \approx \underbrace{[\mathbb{I}_{\mathbf{M}}^R | \mathbb{F}_{\text{bg},\mathbf{N}}]_{\partial V} + [\mathbb{I}_{\mathbf{M}}^R | \mathbb{F}_{\text{scat},\mathbf{N}}]_{\partial V}}_{S_{\mathbf{MN}}} + \underbrace{[\mathbb{I}_{\mathbf{M}}^R | \mathbb{F}_{\text{scat},\mathbf{N}}^{(1)}]_{\partial V}}_{\delta S_{\mathbf{MN}}}. \tag{24}$$

The first two terms on the right-hand side represent the already known scattering matrix $S_{\mathbf{MN}}$ of the unperturbed resonator, while the third term represents the change $\delta S_{\mathbf{MN}}$ of the scattering matrix due to the perturbation. Let us now evaluate this term. In analogy to Eq. (16), we convert the surface integral into volume integrals:

$$[\mathbb{F}_{\mathbf{M}}^{R}|\mathbb{F}_{\text{scat},\mathbf{N}}^{(1)}]_{\partial V} = [\mathbb{F}_{\text{bg},\mathbf{M}}^{R}|\mathbb{F}_{\text{scat},\mathbf{N}}^{(1)}]_{\partial V} = -\langle \mathbb{F}_{\text{bg},\mathbf{M}}^{R}|\delta\hat{\mathbb{M}}|\mathbb{F}_{\text{bg},\mathbf{N}}\rangle_V - \langle \mathbb{F}_{\text{bg},\mathbf{M}}^{R}|\delta\hat{\mathbb{M}}|\mathbb{F}_{\text{scat}}\rangle_V - \langle \mathbb{F}_{\text{bg},\mathbf{M}}^{R}|\Delta\hat{\mathbb{M}}|\mathbb{F}_{\text{scat}}^{(1)}\rangle_V. \tag{25}$$

Here, we used the relations provided under Eq. (16), together with the fact that $\mathbb{F}_{\text{scat},\mathbf{N}}^{(1)}$ is only composed of outgoing basis functions on the surface $\partial V$, as well as Eq. (22).

We insert Eq. (25) into Eq. (24), plug in the expansions provided in Eqs. (23) and (13), exploit the fact that $\delta\hat{\mathbb{M}}$ is only nonzero inside the volume $V_c$, and use the coefficients $a_{n,\mathbf{M}}$ and $b_{n,\mathbf{N}}$ introduced in Eq. (17). All together, we obtain

$$\delta S_{\mathbf{MN}} = \underbrace{-\langle \mathbb{F}_{\text{bg},\mathbf{M}}^{R}|\delta\hat{\mathbb{M}}|\mathbb{F}_{\text{bg},\mathbf{N}}\rangle_{V_c}}_{\equiv \delta S_{\mathbf{MN}}^{\text{nr}}} + \underbrace{\sum_n \frac{a_{n,\mathbf{M}}\langle \mathbb{F}_n^{R}|\delta\hat{\mathbb{M}}|\mathbb{F}_{\text{bg},\mathbf{N}}\rangle_{V_c}}{k - k_n}}_{\equiv \delta S_{\mathbf{MN}}^{\text{ex}}} + \underbrace{\sum_n \frac{b_{n,\mathbf{N}}\langle \mathbb{F}_{\text{bg},\mathbf{M}}^{R}|\delta\hat{\mathbb{M}}|\mathbb{F}_n\rangle_{V_c}}{k - k_n}}_{\equiv \delta S_{\mathbf{MN}}^{\text{em}}}$$
$$\underbrace{-\sum_n \frac{a_{n,\mathbf{M}}b_{n,\mathbf{N}}\langle \mathbb{F}_n^{R}|\delta\hat{\mathbb{M}}|\mathbb{F}_n\rangle_{V_c}}{(k - k_n)^2}}_{\equiv \delta S_{\mathbf{MN}}^{\text{shift}}} - \underbrace{\sum_{n\neq n'} \frac{a_{n,\mathbf{M}}b_{n',\mathbf{N}}\langle \mathbb{F}_n^{R}|\delta\hat{\mathbb{M}}|\mathbb{F}_{n'}\rangle_{V_c}}{(k - k_n)(k - k_{n'})}}_{\equiv \delta S_{\mathbf{MN}}^{\text{cross}}}. \tag{26}$$

This equation constitutes the main finding of the paper and allows to predict the change of the scattering matrix under material perturbations in the resonator. We can identify five different contributions: $\delta S^{\text{nr}}$, $\delta S^{\text{ex}}$, $\delta S^{\text{em}}$, $\delta S^{\text{shift}}$, and $\delta S^{\text{cross}}$. The first one contains an overlap integral between the background fields and represents a nonresonant interaction. The second and third one contain overlap integrals of the modes with the background fields, which describe the change in the excitation and emission efficiencies of these modes, respectively. The fourth one contains an overlap integral of the modes with themselves, which is associated with a shift $\Delta k_n$ of the wavenumber eigenvalue, well-known for permittivity and permeability perturbations from previous works as[5,16–21]

$$\Delta k_n = -\langle \mathbb{F}_n^{R}|\delta\hat{\mathbb{M}}(k_n)|\mathbb{F}_n\rangle_{V_c}. \tag{27}$$

The fifth term contains an overlap integral between different modes, which describes intermodal crosstalk. Note that all quantities in Eq. (26) except $\mathbb{F}_n$ and $k_n$ are $k$ dependent.

We want to emphasize that Eqs. (26) and (27) are valid for *any* kind of material perturbation $\delta\hat{\mathbb{M}}$ and are not limited to the chiral material changes that are discussed in the context of this work. Inserting the special $\delta\hat{\mathbb{M}}$, given by Eq. (5) of the main manuscript, that is associated with a change from $\kappa = 0$ to $\kappa \neq 0$, and writing out the integrals $\langle\ldots\rangle_V$ results in Eqs. (6) to (12) of the main manuscript.

## 2 Calculation of the $\Delta$CD signals from the scattering matrices

This section provides the details on how the $\Delta$CD signal and its contributions were calculated from the scattering matrices $S$ and $\delta S$. The $\Delta$CD signal was defined as

$$\Delta\text{CD} = \text{CD}_\kappa - \text{CD}_0, \tag{28}$$

where $\text{CD}_\kappa$ and $\text{CD}_0$ represent the circular dichroism signals of the resonator with and without $\kappa$, respectively. The circular dichroism signals were defined as the absorption difference between left-handed circularly polarized (LCP) and right-handed circularly polarized (RCP) polarized light. The incidence direction was taken from the top. Under this definition, the circular dichroism signals are related to the scattering matrices via

$$\text{CD}_0 = \underbrace{\left(1 - \sum_{\mathbf{M}} |S_{\mathbf{M},\text{LCP top}}|^2\right)}_{\text{Absorption for LCP top input}} - \underbrace{\left(1 - \sum_{\mathbf{M}} |S_{\mathbf{M},\text{RCP top}}|^2\right)}_{\text{Absorption for RCP top input}} \tag{29}$$

and

$$\text{CD}_\kappa = \underbrace{\left(1 - \sum_{\mathbf{M}} |S_{\mathbf{M},\text{LCP top}} + \delta S_{\mathbf{M},\text{LCP top}}|^2\right)}_{\text{Absorption for LCP top input}} - \underbrace{\left(1 - \sum_{\mathbf{M}} |S_{\mathbf{M},\text{RCP top}} + \delta S_{\mathbf{M},\text{RCP top}}|^2\right)}_{\text{Absorption for RCP top input}}, \tag{30}$$

where the sum goes over all energy-carrying output channels $\mathbf{M}$. The contributions of the $\Delta$CD signal were defined as

$$\Delta\text{CD}_{\text{xxx}} = \Delta\text{CD}\big|_{\delta S = \delta S^{\text{xxx}}}, \tag{31}$$

where xxx = {nr, ex, em, shift, cross}.

In the end, let us consider the case $|\delta S_{\mathbf{M,N}}| \ll |S_{\mathbf{M,N}}|$. With very few exceptions, this condition is automatically fulfilled in scenarios, where the first-order perturbation theory is applicable. In particular, this condition holds for all example systems discussed in this work. Under the above assumption, Eqs. (28) to (31) simplify to more intuitive expressions:

$$\Delta \text{CD} \approx \sum_{\mathbf{M}} 2\operatorname{Re}\left(S^*_{\mathbf{M},\text{RCP top}} \delta S_{\mathbf{M},\text{RCP top}}\right) - \sum_{\mathbf{M}} 2\operatorname{Re}\left(S^*_{\mathbf{M},\text{LCP top}} \delta S_{\mathbf{M},\text{LCP top}}\right), \tag{32}$$

$$\Delta \text{CD}_{\text{xxx}} \approx \sum_{\mathbf{M}} 2\operatorname{Re}\left(S^*_{\mathbf{M},\text{RCP top}} \delta S^{\text{xxx}}_{\mathbf{M},\text{RCP top}}\right) - \sum_{\mathbf{M}} 2\operatorname{Re}\left(S^*_{\mathbf{M},\text{LCP top}} \delta S^{\text{xxx}}_{\mathbf{M},\text{LCP top}}\right). \tag{33}$$

# Supplementary References